\documentclass[reprint,aps,prd,superscriptaddress,showkeys,showpacs]{revtex4-1}
\usepackage{epsfig,amsmath,natbib}

\usepackage{aas_macros}
\usepackage{amssymb}
\usepackage{amsmath}
\usepackage{dsfont}
\usepackage{hyperref}
\usepackage{color}
\usepackage{pbox}

\hypersetup{
	colorlinks=false,
	citecolor=green
}

\begin{document}

\title{Emulating the CFHTLenS Weak Lensing data:  Cosmological Constraints from moments and Minkowski functionals}

\author{Andrea Petri}
\email{apetri@phys.columbia.edu}
\affiliation{Department of Physics, Columbia University, New York, NY 10027, USA}
\affiliation{Physics Department, Brookhaven National Laboratory, Upton, NY 11973, USA}

\author{Jia Liu}
\affiliation{Department of Astronomy, Columbia University, New York, NY 10027, USA}

\author{Zolt\'an Haiman}
\affiliation{Department of Astronomy, Columbia University, New York, NY 10027, USA}

\author{Morgan May}
\affiliation{Physics Department, Brookhaven National Laboratory, Upton, NY 11973, USA}

\author{Lam Hui}
\affiliation{Department of Physics, Columbia University, New York, NY 10027, USA}

\author{Jan M. Kratochvil}
\affiliation{Astrophysics and Cosmology Research Unit, University of KwaZulu-Natal, Westville, Durban 4000, South Africa}

\date{\today}

\label{firstpage}

\begin{abstract}
Weak gravitational lensing is a powerful cosmological probe, with
non--Gaussian features potentially containing the majority of the
information. We examine constraints on the parameter triplet
$(\Omega_m,w,\sigma_8)$ from non-Gaussian features of the weak lensing
convergence field, including a set of moments (up to $4^{\rm th}$
order) and Minkowski functionals, using publicly available data from
the 154\,deg$^2$ CFHTLenS survey. We utilize a suite of ray--tracing
N-body simulations spanning 91 points in $(\Omega_m,w,\sigma_8)$
parameter space, replicating the galaxy sky positions, redshifts and
shape noise in the CFHTLenS catalogs. We then build an emulator that
interpolates the simulated descriptors as a function of
$(\Omega_m,w,\sigma_8)$, and use it to compute the likelihood function
and parameter constraints. We employ a principal component analysis to
reduce dimensionality and to help stabilize the constraints with
respect to the number of bins used to construct each statistic. Using
the full set of statistics, we find
$\Sigma_8\equiv\sigma_8(\Omega_m/0.27)^{0.55}=0.75\pm0.04$ (68\%
C.L.), in agreement with previous values. We find that constraints on
the $(\Omega_m,\sigma_8)$ doublet from the Minkowski functionals
suffer a strong bias. However, high-order moments break the
$(\Omega_m,\sigma_8)$ degeneracy and provide a tight constraint on
these parameters with no apparent bias.  The main contribution comes
from quartic moments of derivatives.
\end{abstract}

\keywords{Weak Gravitational Lensing --- Data analysis --- Methods: analytical,numerical,statistical}
\pacs{98.80.-k, 95.36.+x, 95.30.Sf, 98.62.Sb}

\maketitle

\section{Introduction}

Weak gravitational lensing (hereafter WL) is emerging as a promising
technique to constrain cosmology. Techniques have been developed to
construct cosmic shear fields with shape measurements in large galaxy
catalogues. Although the shear two--point function (2PCF) is the most
widely studied cosmological probe (see, e.g. \citep{CFHTKilbinger}),
alternative statistics have been shown to increase the amount of
cosmological information one can extract from weak lensing
fields. Among these, high--order moments
(\citep{moments1,moments2,moments3,moments4,moments5}),
three--point functions (\citep{3pcf1,3pcf2,3pcf3}), bispectra
(\citep{bispectrum1,bispectrum2,bispectrum3,bispectrum4}), peak counts
(\citep{peaks1,peaks2,peaks3,peaks4,peaks5,peaks6,peaks7,peaks8}) and Minkowski
Functionals (\citep{MinkJan,Petri2013}) have been shown to improve
cosmological constraints in weak lensing analyses.

In this work, we use the publicly available CFHTLenS data, consisting
of a catalog of $\approx$4.2 million galaxies, combined with a suite
of ray-tracing simulations in 91 different cosmological models to
derive constraints on the cosmological parameters $\Omega_m$,
$\sigma_8$ and the dark energy (DE) equation of state $w$.  The
statistics we consider in this work are the Minkowski functionals
(MFs) and the low--order moments (LM) of the convergence field.
Cosmological parameter inferences from CFHTLenS have been obtained
using the 2PCF~\citep{CFHTKilbinger}, and a number of authors have
investigated the constraining power of CFHTLenS using statistics that
go beyond the usual quadratic ones. Fu et al.~\citep{CFHTFu} used
three--point correlations as an additional probe for cosmology, and
found modest ($10-20\%$) improvements over the 2PCF. These results rely on the third order statistics systematic tests performed by Simon et al. \citep{CFHTSimon}.  

Liu et al.~\citep{Companion} have found a more significant ($50-60\%$)
tightening of the $\Omega$ and $\sigma_8$ constraints, utilizing the
abundance of WL peaks. Cosmological constraints using WL peaks in CFHTLenS Stripe82 data
have also been investigated by Liu et. al.~\citep{Stripe82}. 
Finally, closest to the present paper, Shirasaki \&
Yoshida~\citep{CFHTMasato} investigated constraints from Minkowski
Functionals, including systematic errors.  Our study represents two
major improvements over previous work.  First, constraints from the
MFs in~ref.~\citep{CFHTMasato} were obtained through the Fisher matrix
formalism, assuming linear dependence on cosmological parameters. Our
study utilizes a suite of simulations sampling the cosmological
parameter space, mapping out the non-linear parameter-dependence of
each descriptor.  Second, we include the LMs as a set of new
descriptors; these yield the tightest and least biased constraints.

This paper is organized as follows: we first give an overview of the
CFHTLenS catalogs, and summarize the adopted data reduction
techniques. Next, we give a description of our simulation pipeline,
including the ray--tracing algorithm, and the procedure used to sample
the parameter space. We call the statistical weak lensing observables
-- the power spectrum, Minkowski functionals, and moments --
"descriptors" throughout the paper. We discuss the calculation of the
descriptors, including dimensional reduction using a principal
component analysis, and the statistical inference framework we
used. We then describe our main results, i.e. the cosmological
parameter constraints. To conclude, we then discuss our findings and
comment on possible future extensions of this analysis.

\section{Data and simulations}

\subsection{CFHTLenS data reduction}
\label{cfhtdatareduction}

In this section, we briefly summarize our treatment of the public
CFHTLenS data.  For a more in--depth description of our data reduction
procedure, we refer the reader to \citep{Companion}.

The CFHTLenS survey covers four sky patches of 64, 23, 44 and 23
deg$^2$ area, for a total of 154 deg$^2$. The publicly released data
consist of a galaxy catalog created using SExtractor
\citep{SExtractor}, and includes photometric redshifts estimated with
a Bayesian photometric redshift code \citep{PhotoCode} and galaxy
shape measurements using \textit{lensfit} \citep{cfht1,cfht2}.

We apply the following cuts to the galaxy catalog: \texttt{mask} $<1$ (see
Table B2 in \citep{SExtractor}), redshift $0.2 < z < 1.3$ (see
\citep{cfht1}), \texttt{fitclass} = 0 (which requires the object to be a
galaxy) and weight $\mathrm{w}>0$ (with larger $\mathrm{w}$ indicating
smaller shear measurement uncertainty). Applying these cuts leaves us
4.2$\times10^6$ galaxies, 124.7 deg$^2$ sky coverage, and average
galaxy density $n_{gal} \approx 9.3\,\mathrm{arcmin}^{-2}$. The
catalog is further reduced by $\sim25\%$ when one rejects fields with
non--negligible star--galaxy correlations. These spurious correlations
are likely due to imperfect PSF removal, and do not contain
cosmological signal. These cuts are consistent with the ones adopted
by the CFHTLenS collaboration (see \citep{CFHTFu}).

The CFHTLenS galaxy catalog provides us with the sky position
$\pmb{\theta}$, redshift $z(\pmb{\theta})$ and ellipticity
$\mathbf{e}(\pmb{\theta})$ of each galaxy, as well as the individual
weight factors $w(\pmb{\theta})$ and additive and multiplicative
ellipticity corrections $c(\pmb{\theta}), m(\pmb{\theta})$. Because
the CFHTLenS fields are irregularly shaped, we first divide them into
13 squares (subfields) to match the shape and $\approx12$ deg$^2$ size
of our simulated maps (see below). These square-shaped subfield maps
are pixelized according to a Gaussian gridding procedure

\begin{equation}
\bar{\mathbf{e}}(\pmb{\theta}) = \frac{\sum_{i=1}^{N_s} W(\vert\pmb{\theta}-\pmb{\theta}_i\vert)\mathrm{w}(\pmb{\theta}_i)[\mathbf{e}^{obs}(\pmb{\theta}_i)-c(\pmb{\theta}_i)]}{\sum_{i=1}^{N_s}W(\vert\pmb{\theta}-\pmb{\theta}_i\vert)\mathrm{w}(\pmb{\theta}_i)[1+m(\pmb{\theta})]},
\end{equation} 
\begin{equation}
\label{gausskernel}
W_{\theta_G}(\pmb{\theta}) = \frac{1}{2\pi\theta_G^2}\exp{\left(-\frac{\pmb{\theta}^2}{2\theta_G^2}\right)},
\end{equation}
where the smoothing scale $\theta_G$ has been fixed at 1.0\,arcmin
(but varied occasionally to 1.8 and 3.5\,arcmin for specific tests
described below) and $m,c$ refer to the multiplicative and additive
corrections of the galaxies in the catalog.

Using the ellipticity grid $\bar{\mathbf{e}}(\pmb{\theta})$ as an
estimator for the cosmic shear $\gamma^{1,2}(\pmb{\theta})$, we
perform a non--local Kaiser--Squires inversion \citep{KS} to recover
the convergence $\kappa(\pmb{\theta})$ from the $E$--mode of the shear
field,
\begin{equation}
\kappa(\mathbf{l}) = \left(\frac{l_1^2-l_2^2}{l_1^2+l_2^2}\right)\gamma^1(\mathbf{l}) + 2\frac{l_1l_2}{l_1^2+l_2^2}\gamma^2(\mathbf{l}).
\end{equation}
The simulated $\kappa$ maps we create below are $12~\mathrm{deg}^2$ in
size and have a resolution of $512\times512$ pixels. The CFHTLenS
catalogs contain masked regions (which include the rejected fields and
the regions around bright stars). We first create gridded versions of
the observed $\kappa$ maps matching the size and pixel resolution of
our simulated maps, with each pixel containing the number of galaxies
($n_{gal}$) falling within its window.  
We then smooth this galaxy surface density map with the same Gaussian
window function as equation (\ref{gausskernel}) and remove regions
where $n_{gal} < 5 \,\mathrm{arcmin}^{−2}$ (see
\citep{CFHTMasato}). Regions with low galaxy number density can induce
large errors in the cosmological parameter inferences. 

\subsection{Simulation design}

We next give a description of our method to sample the parameter space
with a suite of N--body simulations. We wish to investigate the
non--linear dependence of the descriptors (in this work, Minkowski
Functionals and moments of the $\kappa$ field) on the parameter
triplet $\mathbf{p}=(\Omega_m,w,\sigma_8)$, while keeping the other
relevant parameters $(h,\Omega_b,n_s)$ fixed to the values (0.7,
0.046, 0.96) (see \citep{WMAP9}).
We sampled the $D$--dimensional ($D=3$ in this case) parameter space
using an irregularly spaced grid. The grid was designed with a method
similar to that used to construct an emulator for the matter power
spectrum in the \texttt{Coyote} simulation suite \citep{coyote2}.
Given fixed available computing resources, the irregular grid design
is more efficient than a parameter grid with regular spacings: to
achieve the same average spacing between models in the latter approach
would require a prohibitively large number of simulations.

We limit the parameter sampling to a box whose sides range over
$\Omega_m\in[0.07,1],\,w\in[-3.0,0],\,\sigma_8\in[0.1,1.5]$.  These
are large ranges, with most of the corresponding 3D parameter volume
ruled out by other cosmological experiments. However, our focus in
this work is to quantify the constraints from CFHTLenS alone, which,
by itself has strong parameter degeneracies.
We next map this sampling box $\Pi$ into a hypercube of unit side. We
want to construct an irregularly spaced grid consisting of $N$ points
$\mathbf{x}_i\in[0,1]^D$. Let a \textit{design} $\mathcal{D}$ be the
set of this irregularly spaced $N$ points. Our goal is to find an
optimal design, in which the points are spread as uniformly as
possible inside the box. Following ref.~\citep{coyote2}, we choose our
optimal design as the minimum of the cost function

\begin{equation}
\label{costfunction}
\mathcal{C}(\mathcal{D}) = \frac{2D^{1/2}}{N(N-1)}\sum_{i<j}^N\frac{1}{\vert\mathbf{x}_i-\mathbf{x}_j\vert}.
\end{equation} 
This problem is mathematically equivalent to the minimization of the
Coulomb potential energy of $N$ unit charges in a unit box, which
corresponds to spreading the charges as evenly as possible.
Finding the optimal design $\mathcal{D}_m$ that minimizes
(\ref{costfunction}) can be computationally very demanding, and hence
we decided to use a simplified approach. Although approximate, the
following iterative procedure gives satisfactory accuracy for our purposes:

\begin{enumerate}
\item We start from the diagonal design $\mathcal{D}_0$:
$x_i^d\equiv i/(N-1)$ for $d=1...D$.
\item We shuffle the coordinates of the particles in each dimension independently $x_i^d = \mathcal{P}_d\left(\frac{1}{N-1},\frac{2}{N-1},...,1\right)$, where $\mathcal{P}_1,...,\mathcal{P}_D$ are random independent permutations of $(1,2,...,N)$.
\item We pick a random particle pair $(i,j)$ and a random coordinate $d\in\{1,...,D\}$ and swap $x_i^d\leftrightarrow x_j^d$.
\item We compute the new cost function. If the value is less than in the previous step, we keep the exchange, otherwise we revert the coordinate swap.
\item We repeat steps 3 and 4 until the relative cost function change is less than a chosen accuracy parameter $\epsilon$.
\end{enumerate}

\begin{figure*}
\begin{center}
\includegraphics[scale=0.4]{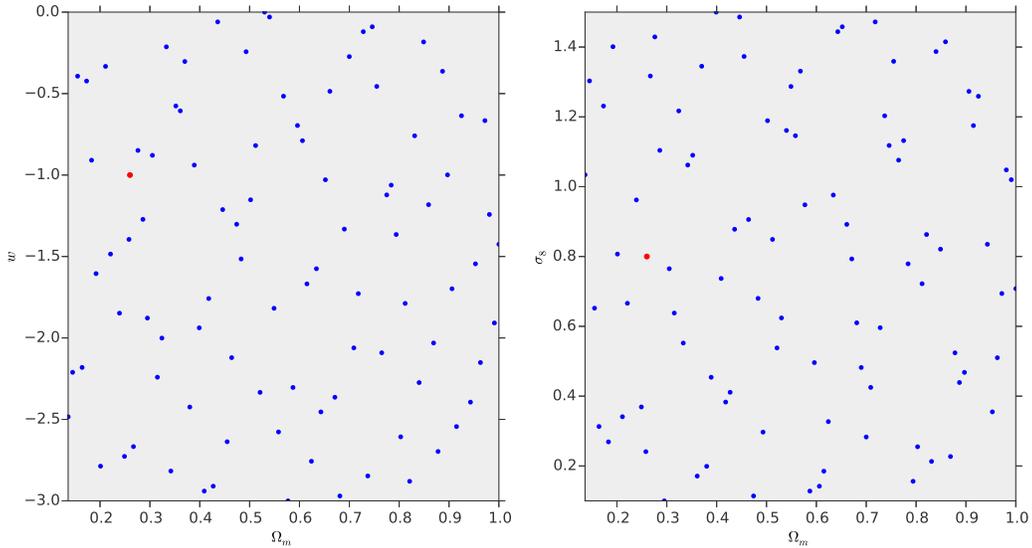}
\caption{$(\Omega_m,w)$ and $(\Omega_m,\sigma_8)$ projections of the final simulation design. The blue points correspond to the \texttt{CFHTemu1} simulation set, which consists of one $N$--body simulation per point, while the red point corresponds to the \texttt{CFHTcov} simulation set, which is based on 50 independent $N$--body simulations.}
\label{designfig}
\end{center}
\end{figure*}

We have found that for $N=91$ grid points O($10^5$) iterations are
sufficient to reach an accuracy of $\epsilon\sim10^{-4}$. Once the
optimal design $\mathcal{D}_m$ has been determined, we invert the
mapping $\Pi\rightarrow[0,1]^3$ to arrive at our simulation parameter
sampling $\mathbf{p}_s$. We show the final list of grid points in
Table~\ref{designtable} and Figure~\ref{designfig}.

\begin{table*}
\begin{tabular}{c|ccc||c|ccc||c|ccc||c|ccc}
$N$ & $\Omega_m$ & $w$ & $\sigma_8$ & $N$ & $\Omega_m$ & $w$ & $\sigma_8$ & $N$ & $\Omega_m$ & $w$ & $\sigma_8$ & $N$ & $\Omega_m$ & $w$ & $\sigma_8$ \\ \hline
1 & 0.136 & -2.484 & 1.034 & 26 & 0.380 & -2.424 & 0.199 & 51 & 0.615 & -1.668 & 0.185 & 76 & 0.849 & -0.183 & 0.821 \\
2 & 0.145 & -2.211 & 1.303 & 27 & 0.389 & -0.939 & 0.454 & 52 & 0.624 & -2.757 & 0.327 & 77 & 0.859 & -1.182 & 1.415 \\
3 & 0.155 & -0.393 & 0.652 & 28 & 0.399 & -1.938 & 1.500 & 53 & 0.634 & -1.575 & 0.976 & 78 & 0.869 & -2.031 & 0.227 \\
4 & 0.164 & -2.181 & 0.313 & 29 & 0.409 & -2.940 & 0.737 & 54 & 0.643 & -2.454 & 1.444 & 79 & 0.878 & -2.697 & 0.524 \\
5 & 0.173 & -0.423 & 1.231 & 30 & 0.418 & -1.758 & 0.383 & 55 & 0.652 & -1.029 & 1.458 & 80 & 0.887 & -0.363 & 0.439 \\
6 & 0.183 & -0.909 & 0.269 & 31 & 0.427 & -2.910 & 0.411 & 56 & 0.661 & -0.486 & 0.892 & 81 & 0.897 & -0.999 & 0.468 \\
7 & 0.192 & -1.605 & 1.401 & 32 & 0.436 & -0.060 & 0.878 & 57 & 0.671 & -2.364 & 0.793 & 82 & 0.906 & -1.698 & 1.273 \\
8 & 0.201 & -2.787 & 0.807 & 33 & 0.446 & -1.212 & 1.486 & 58 & 0.681 & -2.970 & 0.610 & 83 & 0.915 & -2.544 & 1.175 \\
9 & 0.211 & -0.333 & 0.341 & 34 & 0.455 & -2.637 & 1.373 & 59 & 0.690 & -1.332 & 0.482 & 84 & 0.925 & -0.636 & 1.259 \\
10 & 0.221 & -1.485 & 0.666 & 35 & 0.464 & -2.121 & 0.906 & 60 & 0.700 & -0.273 & 0.283 & 85 & 0.943 & -2.394 & 0.835 \\
11 & 0.239 & -1.848 & 0.962 & 36 & 0.474 & -1.302 & 0.114 & 61 & 0.709 & -2.061 & 0.425 & 86 & 0.953 & -1.545 & 0.355 \\
12 & 0.249 & -2.727 & 0.369 & 37 & 0.483 & -1.515 & 0.680 & 62 & 0.718 & -1.728 & 1.472 & 87 & 0.963 & -2.151 & 0.510 \\
13 & 0.258 & -1.395 & 0.241 & 38 & 0.493 & -0.243 & 0.297 & 63 & 0.728 & -0.120 & 0.596 & 88 & 0.972 & -0.666 & 0.694 \\
14 & 0.267 & -2.667 & 1.317 & 39 & 0.502 & -1.152 & 1.189 & 64 & 0.737 & -2.847 & 1.203 & 89 & 0.981 & -1.242 & 1.048 \\
15 & 0.276 & -0.849 & 1.429 & 40 & 0.512 & -0.819 & 0.849 & 65 & 0.746 & -0.090 & 1.118 & 90 & 0.991 & -1.908 & 1.020 \\
16 & 0.286 & -1.272 & 1.104 & 41 & 0.521 & -2.334 & 0.538 & 66 & 0.755 & -0.456 & 1.359 & 91 & 1.000 & -1.425 & 0.708 \\
17 & 0.295 & -1.878 & 0.100 & 42 & 0.530 & 0.000 & 0.624 & 67 & 0.765 & -2.091 & 1.076 & -- & -- & -- & -- \\
18 & 0.305 & -0.879 & 0.765 & 43 & 0.540 & -0.030 & 1.161 & 68 & 0.775 & -1.122 & 1.132 & -- & -- & -- & -- \\
19 & 0.315 & -2.241 & 0.638 & 44 & 0.549 & -1.818 & 1.287 & 69 & 0.784 & -1.062 & 0.779 & -- & -- & -- & -- \\
20 & 0.324 & -2.001 & 1.217 & 45 & 0.558 & -2.577 & 1.146 & 70 & 0.794 & -1.365 & 0.156 & -- & -- & -- & -- \\
21 & 0.333 & -0.213 & 0.552 & 46 & 0.568 & -0.516 & 1.331 & 71 & 0.803 & -2.607 & 0.255 & -- & -- & -- & -- \\
22 & 0.342 & -2.817 & 1.062 & 47 & 0.577 & -3.000 & 0.948 & 72 & 0.812 & -1.788 & 0.722 & -- & -- & -- & -- \\
23 & 0.352 & -0.576 & 1.090 & 48 & 0.587 & -2.304 & 0.128 & 73 & 0.821 & -2.880 & 0.863 & -- & -- & -- & -- \\
24 & 0.361 & -0.606 & 0.171 & 49 & 0.596 & -0.696 & 0.496 & 74 & 0.831 & -0.759 & 0.213 & -- & -- & -- & -- \\
25 & 0.370 & -0.303 & 1.345 & 50 & 0.606 & -0.789 & 0.142 & 75 & 0.840 & -2.274 & 1.387 & -- & -- & -- & -- \\
\end{tabular}
\caption{List of the \texttt{CFHTemu1} grid points in the 3D cosmological parameter space.}
\label{designtable}
\end{table*}

For each parameter point on the grid $\mathbf{p}_s$ we then run an
$N$--body simulation and perform ray tracing, as described in
\S~\ref{raysim}, to simulate CFHTLenS shear catalogs.  Throughout the
rest of this paper, we refer to this set of simulations as
\texttt{CFHTemu1}.  Additionally, we have run 50 independent $N$--body
simulations with a \textit{fiducial} parameter choice
$\mathbf{p}_0=(0.26,-1.0,0.8)$, for the purpose of accurately
measuring the covariance matrices, needed for the parameter inferences
in \S\ref{cosmostats}. This additional suite of 50 simulations will
be referred to as \texttt{CFHTcov}.

\subsection{Ray--Tracing Simulations}
\label{raysim}

The goal of this section is to outline our simulation pipeline.  The
fluctuations in the matter density field between a source at redshift
$z$ and an observer located on Earth will cause small deflections in
the trajectories of light rays traveling from the source to the
observer. 
We estimate the dark matter gravitational potential running $N$--body
simulations with $N=512^3$ particles, using the public code Gadget2
\citep{Gadget2}. We adopted a comoving box size of $240h^{-1}$Mpc,
corresponding to a mass resolution of $7.4\times10^9h^{-1}M_\odot$.
The simulations include dark matter only, and the initial conditions
were generated with \texttt{N-GenIC} at $z=100$, based on the linear
matter power spectrum created with the Einstein-Boltzmann code CAMB
\cite{CAMB}. Data cubes were output at redshift intervals
corresponding to $80h^{-1}$ (comoving) Mpc.

Using a procedure similar to
refs.~\citep{RayTracingJain,RayTracingHartlap}, the equation that
governs the light ray deflections can be written in the form
\begin{equation}
\label{raytrajectory}
\frac{d^2\mathbf{x}(\chi)}{d\chi^2} = -\frac{2}{c^2}\nabla_{\mathbf{x}_\perp}\Phi(\mathbf{x}_\perp(\chi),\chi),
\end{equation}
where $\chi$ is the radial comoving distance,
$\mathbf{x}_\perp=\chi\pmb{\beta}$ refers to two transverse
coordinates (with $\pmb{\beta}$ the angular sky coordinates, using the
flat sky approximation)a, $\mathbf{x}(\chi)$ is the trajectory of a
single light ray, and $\Phi$ is the gravitational potential.

Suppose that a light ray reaches the observer at an angular position
$\pmb{\theta}$ on the sky: we want to know where this light ray
originated, knowing it comes from a redshift $z_s$. To answer this
question we need to integrate equation (\ref{raytrajectory}) with the
initial condition $\pmb{\beta}(0;\pmb{\theta})=\pmb{\theta}$ up to a distance
$\chi_s=\chi(z_s)$ to obtain the source angular position
$\pmb{\beta}(\chi_s;\pmb{\theta})$. Since light rays travel undeflected from the observer to the first lens plane, the derivative initial condition in the Cauchy problem (\ref{raytrajectory}) reads $\dot{\pmb{\beta}}(0;\pmb{\theta})=0$. We indicate the derivative of $\pmb{\beta}$ with respect to $\chi$ as $\dot{\pmb{\beta}}$. 
We use our proprietary implementation Inspector Gadget~(see, e.g., \cite{KHM2010}) to solve
for the light ray trajectories based on a discretized version of equation (\ref{raytrajectory}) that is based on the multi--lens--plane algorithm (see \citep{RayTracingHartlap} for example).
Applying random periodical shifts and rotations to the $N$--body simulation data cubes, we generate $R=1000$ pseudo--independent realizations of the lens plane system used to solve (\ref{raytrajectory}). 
Once we obtain the light ray trajectories, we infer the relevant weak
lensing quantities by taking angular derivatives of the ray
deflections $A(\chi_s;\pmb{\theta}) = \partial
\pmb{\beta}(\chi_s;\pmb{\theta})/\partial\pmb{\theta}$ and performing
the usual spin decomposition to infer the convergence $\kappa$ and the
shear components $(\gamma^1,\gamma^2)$,
\begin{equation}
A(\chi_s;\pmb{\theta}) = (1-\kappa(\chi_s;\pmb{\theta}))\pmb{I} - \gamma^1(\chi_s;\pmb{\theta})\sigma^3 - \gamma^2(\chi_s;\pmb{\theta})\sigma^1
\end{equation}  
where $\pmb{I}$ is the $2\times2$ identity and $\sigma^{1,3}$ are the
first and third Pauli matrices. We perform this procedure for each of the $R$ realizations of the lens planes and we obtain $R$ pseudo--independent realizations of the $\pmb{\gamma}$ weak lensing field. We use these different random realizations to estimate the means (from the \texttt{CFHTemu1} simulations) and covariance matrices (from the \texttt{CFHTcov} simulations) of our descriptors. Since the random box rotations and translations that make up the \texttt{CFHTcov} simulations are based on 50 independent $N$--body runs, we believe the covariance matrices measured from this set to be more accurate than the ones measured from the \texttt{CFHTemu1} set. 

The convergence $\kappa$ is related to
the magnification, while the two components of the complex shear
$\pmb{\gamma}=\gamma^1 + i\gamma^2$ are related to the apparent
ellipticity of the source. Given a source with intrinsic complex
ellipticity $\mathbf{e}_s=e^1_s + ie^2_s$, its observed ellipticity
will be modified to
\begin{equation}
\mathbf{e} = 
\begin{cases}
\frac{\mathbf{e}_s+\mathbf{g}}{1+\mathbf{g}^*\mathbf{e}_s} \,\,\,\,\,\,\,\, \vert \mathbf{g}\vert \leq 1 \\ \\
\frac{1+\mathbf{ge}_s^*}{\mathbf{e}_s^* + \mathbf{g}^*} \,\,\,\,\,\,\,\, \vert \mathbf{g}\vert > 1
\end{cases}
\end{equation}
where $\mathbf{g} \equiv \pmb{\gamma}/(1-\kappa)$ is the reduced shear. 

For each simulated galaxy, we assign an intrinsic ellipticity by
rotating the observed ellipticity for that galaxy by a random angle on
the sky, while conserving its magnitude $\vert\mathbf{e}\vert$. To be
consistent with the CFHTLenS analysis, we adopt the weak lensing limit
($\vert\pmb{\gamma}\vert\ll1,\kappa\ll1$), i.e.
$\mathbf{g}\approx\pmb{\gamma}$ and $\mathbf{e}\approx
\mathbf{e}_s+\pmb{\gamma}$. We also add the multiplicative shear
corrections by replacing $\pmb{\gamma}$ with $(1+m)\pmb{\gamma}$. We
note that the observed ellipticity for a particular galaxy already
contains the lensing shear by large scale structure (LSS), but the random
rotation makes this contribution at least second order in $\kappa$ by destroying the shape spatial correlations induced by lensing from LSS. Consistent with the weak lensing approximation, the lensing signal from the
simulations is first order in $\kappa$ and hence the randomly rotated
observed ellipticities can be safely considered as intrinsic
ellipticities. 

We analyze the simulations in the same way as we analyzed the CFHTLenS
data -- constructing the simulated $\kappa$ maps as explained in
\S\ref{cfhtdatareduction}. These final simulation products are then
processed together with the $\kappa$ maps obtained from the data to
compute confidence intervals on the parameter triplet
$(\Omega_m,w,\sigma_8)$.

\section{Statistical methods}

The goal of this section is to describe the framework to combine the
CFHT data and our simulations, and to derive the constraints on the
cosmological parameter triplet $(\Omega_m,w,\sigma_8)$. Briefly, we
measure the same set of statistical descriptors from the data and from the
simulations; these are then compared in a Bayesian framework in
order to compute parameter confidence intervals.

\subsection{Descriptors}

The statistical descriptors we consider in this work are the Minkowski
Functionals (MFs) and the low--order moments (LMs) of the convergence
field. The three MFs $(V_0,V_1,V_2)$ are topological descriptors of
the convergence field $\kappa(\pmb{\theta})$, probing the area,
perimeter and genus characteristic of the $\kappa$ excursion sets
$\Sigma_{\kappa_0}$, defined as
$\Sigma_{\kappa_0}=\{\kappa>\kappa_0\}$. Following
refs.~\citep{Petri2013,MinkJan} we use the following local estimators
to measure the MFs from the $\kappa$ maps:
\begin{equation*}
\label{v0meas}
V_0(\kappa_0)=\frac{1}{A}\int_A\Theta(\kappa(\pmb{\theta})-\kappa_0)d\pmb{\theta},
\end{equation*}
\begin{equation}
\label{v1meas}
V_1(\kappa_0)=\frac{1}{4A}\int_A\delta_D(\kappa(\pmb{\theta})-\kappa_0)\sqrt{\kappa_x^2+\kappa_y^2}d\pmb{\theta},
\end{equation}
\begin{equation*}
\label{v2meas}
V_2(\kappa_0)=\frac{1}{2\pi A}\int_A\delta_D(\kappa(\pmb{\theta})-\kappa_0)\frac{2\kappa_x\kappa_y\kappa_{xy}-\kappa_x^2\kappa_{yy}-\kappa_y^2\kappa_{xx}}{\kappa_x^2+\kappa_y^2}d\pmb{\theta}.
\end{equation*}
Here $A$ is the total area of the field of view and $\kappa_{x,y}$
denotes gradients of the $\kappa$ field, which we evaluate using
finite differences. In this notation $\Theta(x)$ is the Heaviside function and $\delta_D(x)$ is the Dirac delta function.
The first Minkowski functional, $V_0$, is
equivalent to the cumulative one--point PDF of the $\kappa$ field,
while $V_1,V_2$ are sensitive to the correlations between nearby
pixels. The one--point PDF of the $\kappa$ field, $\partial V_0$, can
be obtained by differentiation $\partial
V_0(\kappa_0)=dV_0(\kappa_0)/d\kappa_0$. 

In addition to these topological descriptors, we consider a set of
low--order moments of the convergence field (two quadratic, three
cubic and four quartic). We choose these moments to be the minimal set
of LMs necessary to build a perturbative expansion of the MFs up to
$O(\sigma_0^2)$ (see \citep{Munshi12,Matsubara10}). We adopt the
following definitions
\begin{equation}
\label{momentestimator}
\begin{matrix}
\mathrm{LM_2}: \sigma_{0,1}^2 = \langle\kappa^2\rangle,\langle\vert\nabla\kappa\vert^2\rangle, \\ \\
\mathrm{LM_3}: S_{0,1,2} = \langle\kappa^3\rangle,\langle\kappa\vert\nabla\kappa\vert^2\rangle,\langle\kappa^2\nabla^2\kappa\rangle, \\ \\
\mathrm{LM_4}: K_{0,1,2,3} = \langle\kappa^4\rangle,\langle\kappa^2\vert\nabla\kappa\vert^2\rangle,\langle\kappa^3\nabla^2\kappa\rangle,\langle\vert\nabla\kappa\vert^4\rangle.
\end{matrix}
\end{equation}

If the $\kappa$ field were Gaussian, one could express the MFs in
terms of the LM$_2$ moments, which are the only independent moments
for a Gaussian random field. In reality, weak lensing convergence
fields are non--Gaussian and the MF and LM descriptors are not
guaranteed to be equivalent. Refs.~\citep{Munshi12,Matsubara10}
studied a perturbative expansion of the MFs in powers of the standard
deviation $\sigma_0$ of the $\kappa$ field. When truncated at order
$O(\sigma_0^2)$, this can be expressed completely in terms of the LMs
up to quartic order. Such perturbative series, however, have been
shown not to converge \citep{Petri2013} unless the weak lensing fields
are smoothed with windows of size $\geq 15^\prime$. Because of this,
throughout this work, we treat MF and LM as separate statistical
descriptors.

We note that this choice is somewhat ad-hoc. In general, the LMs that
contain gradients are sensitive to different shapes of the $\kappa$
multispectra $P_{\kappa}^n(\mathbf{l}_1,...,\mathbf{l}_n)$ because a
particular $\mathrm{LM}_n$ has the general form
\begin{equation}
\mathrm{LM}_n = \int d\mathbf{l}_1...d\mathbf{l}_n \rho(\mathbf{l}_{1...n})P^n_{\kappa}(\mathbf{l}_{1...n})
\end{equation}
where $\rho$ is a polynomial of order $n$ in the $\mathbf{l}$'s. For
example for $K_2$ we have $\rho(\mathbf{l}_{1234})=l_4^2$ and this
moment emphasizes quadrilateral shapes for which one side is much
larger than the others. On the other hand, for $K_3$ we have
$\rho(\mathbf{l}_{1234})=(\mathbf{l}_1\cdot\mathbf{l}_2)(\mathbf{l}_3\cdot\mathbf{l}_4)$
and this moment is most sensitive to trispectrum shapes that are close
to rectangular. There are moments which include derivatives in
addition to those included in Eq.~\ref{momentestimator}. In the
future, we will investigate whether there is additional constraining
power in these additional quartic moments.

In addition to the MFs and LMs, we consider the angular power spectrum
$P_l\equiv P^2_l$ of $\kappa$, defined as
\begin{equation}
\label{powerspectrum}
\langle\tilde{\kappa}(\mathbf{l})\tilde{\kappa}(\mathbf{l}')\rangle=(2\pi)^2\delta_D(\mathbf{l}+\mathbf{l}')P_l,
\end{equation}  
where $\tilde{\kappa}(\mathbf{l})$ is the Fourier transform of the
$\kappa$ field. Previous works have studied cosmological constraints from
the convergence power spectrum extensively.  Here our purpose is to
compare the constraints we obtain from the MFs and LMs to ones present
in the literature, which are based on the use of quadratic statistics
(see for example \citep{CFHTKilbinger}). The statistical descriptors
used in this work are summarized in Table \ref{desctable}.
\begin{table}
\begin{tabular}{c|c|c} \hline
Descriptor & Details & $N_b$ (linear spacing) \\ \hline
$V_0,V_1,V_2$ (MF) & $\kappa_0\in[-0.04,0.12]$ & 50 \\
Power Spectrum (PS) & $l \in [300,5000]$ & 50 \\
Moments (LM) & -- & 9 \\
\end{tabular}
\caption{Summary of the descriptors we used, together with the specifications and the number of bins $N_b$ in each case.}
\label{desctable}
\end{table}

When measuring statistical descriptors on $\kappa$ maps, particular
attention must be paid to the effect of masked pixels. The MFs and LMs
remain well--defined in the presence of masks, since the estimators in
equations~(\ref{v1meas}) and (\ref{momentestimator}) are defined
locally, and can be computed in the non--masked regions (with the
exception of the few pixels that are close to the mask
boundaries). The situation is more complicated for power spectrum
measurements, which require the evaluations of Fourier transforms and
hence rely on the value of every pixel in the map. Although
sophisticated schemes to interpolate over the masked regions have been
studied (see for example \citep{VplasInterpolation}), for the sake of
simplicity, we here insert the value $\kappa=0$ in each masked pixel.
Given the uniform spatial distribution of the masked regions in the data, we expect that masks have a little effect on the power spectrum at the range of
multipoles in Table \ref{desctable}, except for an overall normalization which will be the same both in the data and the simulations.  
Likewise, we believe that the way we deal with masked sky regions --
essentially ignoring them -- is robust for the MFs and LMs. Since we
apply the same masks to our simulations and the data, they are
unlikely to introduce biases in the resulting constraints.  Masks, of
course, can still affect the sensitivity and weaken constraints. The
impact of the masks and their treatment has been evaluated for the
MFs, obtained from CFHTLenS, by ref.~\citep{CFHTMasato}, in which the authors find that the masked regions 
are not a dominant source of systematic effects in the CFHTLenS data.

\subsection{Cosmological parameter inferences}
\label{cosmostats}

In this section, we briefly outline the statistical framework
adopted for computing cosmological parameter confidence levels.  We
make use of the MFs and LMs, as well as the power spectrum, as
discussed in the previous section. We refer to $M_i^r(\mathbf{p})$ as
the descriptor measured from a realization $r$ of one of our
simulations with a choice of cosmological parameters $\mathbf{p}$
(i.e. from one of the $R=1000$ map realizations in this cosmology), and to $D_i$
as the descriptor measured from the CFHTLenS data.
In this notation, $i$ is an index that refers to the particular bin on
which the descriptor is evaluated (for example $i$ can range from 0 to
9 for the LM statistic and from 0 to $N_b-1$ for a MF measured in
$N_b$ different, linearly spaced $\kappa$ bins, as indicated in Table \ref{desctable}).

Once we make an assumption for the data likelihood
$\mathcal{L}_d(D_i\vert \mathbf{p})$ and for the parameter priors
$\Pi(\mathbf{p})$, we can use Bayes' theorem to compute the parameter
likelihood $\mathcal{L}_p$,

\begin{equation}
\label{parameterlikelihood}
\mathcal{L}_p(\mathbf{p}\vert D_i) = \frac{\mathcal{L}_d(D_i\vert \mathbf{p})\Pi(\mathbf{p})}{N_{\mathcal{L}}}.
\end{equation}
Here $N_{\mathcal{L}}$ is a $\mathbf{p}$--independent constant that
ensures the proper normalization for $\mathcal{L}_p$. We make the
usual assumption that the data likelihood $\mathcal{L}_d(D_i\vert
\mathbf{p})$ is Gaussian~\footnote{In principle, this assumption could
  be relaxed, by measuring joint probability distributions of the
  descriptors in our simulations.  In practice, characterizing
  non-Gaussianities and folding them into the likelihood analysis
  requires significant analysis, which we postpone to future work.}

\begin{equation}
\label{datalikelihood}
\begin{matrix}
\mathcal{L}_d(D_i\vert \mathbf{p}) = [(2\pi)^{N_b}\det{\mathbf{C}}]^{-1/2} e^{-\frac{1}{2}\chi^2(D_i\vert \mathbf{p})}, \\ \\
\chi^2(D_i\vert \mathbf{p}) = \mathbf{[D - M(p)]}^T\mathbf{C^{-1}[D-M(p)]}.
\end{matrix}
\end{equation} 
We assume, for simplicity, that the covariance matrix $\mathbf{C}$ in
equation~(\ref{datalikelihood}) is $\mathbf{p}$--independent and
coincides with $\mathbf{C}(\mathbf{p}_0)$. The simulated descriptors $\mathbf{M(p)}$ are
measured from an average over the $R=1000$ realizations in the \texttt{CFHTemu1} ensemble  
\begin{equation}
M_i(\mathbf{p}) = \frac{1}{R}\sum_{r=1}^R M_i^r .
\end{equation}
The covariance matrix 
\begin{equation}
\label{datacov}
C_{ij} = \frac{1}{R-1} \sum_{r=1}^R [M_i^r(\mathbf{p}_0)-M_i(\mathbf{p}_0)][M_j^r(\mathbf{p}_0)-M_j(\mathbf{p}_0)]
\end{equation}
is measured from the  $R=1000$ realizations in the \texttt{CFHTcov} ensemble.
While eq.~(\ref{datacov}) gives an unbiased estimator of the
covariance matrix, its inverse is not an unbiased estimator of
$\mathbf{C}^{-1}$ (e.g. ref.~\citep{RayTracingHartlap}). Given that in
our case $R\gg N_b$, we can safely neglect the correction factor
needed to make the estimator for $\mathbf{C}^{-1}$ unbiased.

When computing parameter constraints from the CFHTLenS weak lensing
data alone, we make a flat prior assumption for $\Pi(\mathbf{p})$.  We
postpone using different priors, incorporating external data, for
future work. Parameter inferences are made estimating the location of
the maximum of the parameter likelihood in
eq.~(\ref{parameterlikelihood}), which we call $\mathbf{p}_{ML}(D_i)$,
as well as its confidence contours. The $N\sigma$--confidence contour
of $\mathcal{L}_p(\mathbf{p}\vert D_i)$ is defined to be the subset of
points in parameter space on which the likelihood has a constant value
$c_N$ and
\begin{equation}
\label{ennesigma}
\int_{\mathcal{L}>c_N} \mathcal{L}_p(\mathbf{p}\vert D_i) d\mathbf{p} = \frac{1}{\sqrt{2\pi}}\int_{-N}^N dx e^{-x^2/2}.
\end{equation}

Using equation (\ref{rbfinterpolation}) below, and given the low
dimensionality of the parameter space we consider $(D=3)$, we are able
to directly compute the parameter likelihood
eq.~(\ref{parameterlikelihood}) for $100^3$ different combinations of
the cosmological parameters $p$, arranged in a finely spaced
$100\times100\times100$ mesh within the prior window
$\Pi(\mathbf{p})$. We directly compute the maximum likelihood
$\mathbf{p}_{ML}(D_i)$ and the contour levels $c_N$ without the need
for more sophisticated MCMC methods.

The data likelihood is directly available for parameter combinations
on the simulated irregular grid $\mathbf{p}_s$.  We use a Radial Basis
Function (RBF) scheme to interpolate $M(\mathbf{p})$ to arbitrary
intermediate points.  We approximate the model descriptor as
\begin{equation}
\label{rbfinterpolation}
M(\mathbf{p}) = \sum_{s=1}^N \lambda_s\phi(\vert\mathbf{p}-\mathbf{p}_s\vert)
\end{equation}
where $\phi$ has been chosen as a multi-quadric function
$\phi(r)=\sqrt{1+(r/r_0)^2}$, with $r_0$ chosen as the mean Euclidean
distance between the points in the simulated grid $\mathbf{p}_s$. The
constant coefficients $\lambda_s$ can be determined by imposing the
$N$ constraints $M(\mathbf{p}=\mathbf{p}_s)=M(\mathbf{p_s})$, which
enforce exact results at the simulated points. The interpolation
computations are conveniently performed using the
\texttt{interpolate.Rbf} routine contained in a Scipy library~\citep{scipy}.

\begin{figure*}
\begin{center}
\includegraphics[scale=0.4]{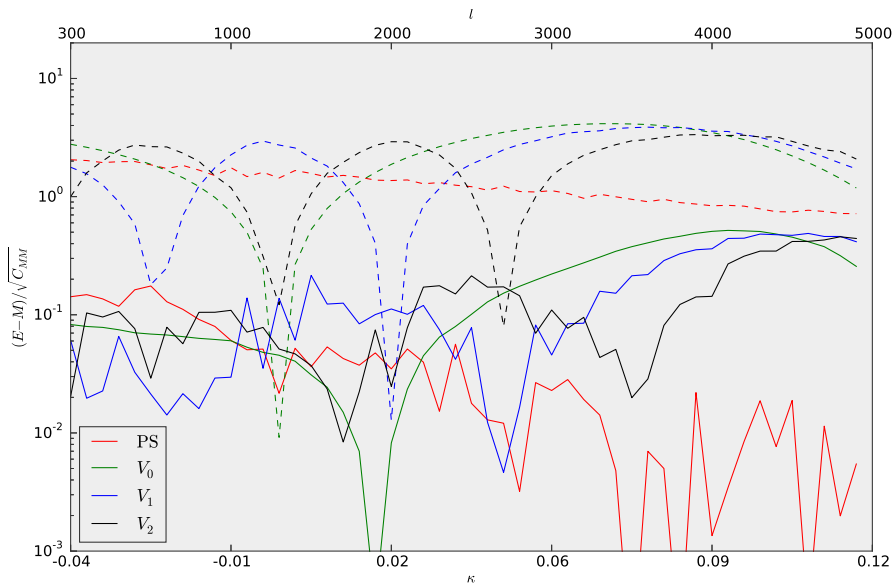}
\end{center}
\caption{Accuracy of the emulator based on the \texttt{CFHTemu1}
  simulations. The figure shows the absolute difference between the descriptor
  interpolated at the fiducial parameter setting, and the descriptor
  expected from the \texttt{CFHTcov} simulations (these are the absoulte values of differences which oscillate around zero). The descriptors are
  shown in units of the standard deviation in each bin $i$ (determined
  from the diagonal elements of the \texttt{CFHTcov} covariance
  matrix). We show the accuracy results for the power spectrum (red)
  and the three Minkowski functionals $V_0$ (green), $V_1$ (blue) and
  $V_2$ (black). For reference, we also show, using dashed lines, the
  difference between the expected \texttt{CFHTcov} descriptors and the
  interpolated descriptor at the non--fiducial point $\mathbf{p}=(0.8,-1.0,0.5)$. This non--fiducial point lies beyond the LM $1\sigma$ contour from the simulations shown in Figure \ref{contours3single} right panel, and corresponds to the target accuracy we wish to achieve}
\label{emulatorAccuracy}
\end{figure*}

We studied the accuracy of the emulator, built with the
\texttt{CFHTemu1} simulations, by interpolating the convergence
descriptors to the fiducial parameter setting
$(\Omega_m,w,\sigma_8)=(0.26,-1.0,0.8)$ and comparing the result to
the one expected from the \texttt{CFHTcov}
simulations. Figure~\ref{emulatorAccuracy} shows that our power
spectrum emulator has a relative error smaller than 20\% for the lower
multipoles ($l<500$), and comparable to 1\% for the higher multipoles. The MF
emulator has a relative error $\lesssim$10\% for the first 30 bins (which
correspond to $\kappa$ values in $[-0.04,0.08]$) and deteriorates due
to numerical noise for the remaining 20 bins. We eliminate the
residual impact of these inaccuracies using a dimensionality reduction
framework, which we explain in the next section. Nevertheless, we do
not expect these inaccuracies to affect our
conclusions. Figure~\ref{emulatorAccuracy} demonstrates that our
emulator is able to distinguish a non--fiducial model from the
fiducial one within numerical errors.  We thus found no need to
implement a more sophisticated interpolation
scheme~\footnote{Following ref.~\citep{coyote2}, we have experimented with interpolations using a Gaussian Process, but did not find a significant improvement in relative error over the much simpler polynomial scheme; this is in agreement with earlier results (see for example \citep{KnoxGP})}.

\subsection{Dimensionality reduction}
\label{pcasection}

The main goal of this work is constraining the cosmological parameter
triplet $(\Omega_m,w,\sigma_8)$ using the CFHTLenS data.  Once the
$N\sigma$ contours have been obtained, using the procedure and
equations (\ref{parameterlikelihood})--(\ref{ennesigma}) outlined
above, one may ask whether the choice of binning affects these
contours.  Indeed, in our previous work, we have found that the number
of bins, $N_b$, can have a non-negligible effect on the contour sizes
(see \citep{Petri2013} for an example with simulated datasets).

In order to ensure that are results are robust with respect to binning
choices, we have implemented a Principal Component Analysis (PCA)
approach. Our physical motivation for this approach is that, even
though we need to specify $N_b$ numbers in order to fully characterize
a binned descriptor, we suspect that the majority of the constraining
information (of a particular descriptor) is contained in a limited
number of linear combinations of its binning. In the framework adopted
by \citep{coyote2}, for example, the authors find that the majority of
the cosmological information in the matter power spectrum is contained
in only 5 different linear combinations of the multipoles. Because of
this, we believe that dimensionality reduction techniques such as PCA
can help deliver accurate cosmological constraints using only a
limited number of descriptor degrees of freedom.

In order to compute the principal components of our descriptor space,
we use the \texttt{CFHTemu1} simulations, which sample the
cosmological parameters at the $N=91$ points listed in Table
\ref{designtable}, and allow us to compute the $N\times N_b$ model
matrix $M_{pi}=M_i(\mathbf{p})$. Note that this is a rectangular
(non-square) matrix. Following a standard procedure (see, e.g.,
ref.~\citep{astroMLText}), we derive the {\it whitened} model matrix
$\tilde{M}_{pi}$, defined by subtracting the mean (over the $N=91$ models) of each bin,
and normalizing it by its variance (always over the $N=91$ models). Next we proceed with a singular
value decomposition (SVD) of $\mathbf{\tilde{M}}$,
\begin{equation}
\label{svd}
\mathbf{U}\mathbf{S} \mathbf{V}^T=\frac{\mathbf{\tilde{M}}}{\sqrt{N-1}},
\end{equation}   
where $S_{ij}=S_i\delta_{ij}$ is a diagonal matrix and $V^T_{ij}$ is
the $j$--th coordinate ($j=1...N_b$) of the $i$--th principal
component ($i=1...\mathrm{min}[N_b,N]$) of $\mathbf{\tilde{M}}$, with
the index $j$ ranging from $1$ to $N_b$. By construction,
$\mathbf{V}$ is $\mathbf{p}$--independent.

To rank the Principal Components $V^T$ in order of importance, we note
that the diagonal matrix $\mathbf{S}^2$ is simply the diagonalization
of the model covariance (not to be confused with the descriptor
covariance in eq.~(\ref{datacov})),
\begin{equation}
\frac{1}{N-1}\mathbf{\tilde{M}}^T\mathbf{\tilde{M}} = \mathbf{V}\mathbf{S}^2\mathbf{V}^T.
\end{equation} 
We follow the standard interpretation of PCA components, stating that
the only meaningful components $V^T_i$ in the analysis (i.e. the ones
that contain the relevant cosmological information) are those
corresponding to the largest eigenvalues $S^2_{i}$, with the smallest
eigenvalues corresponding to noise in the model, due to numerical
inaccuracies in the simulation pipeline. We expect our constraints to
be stable with respect to the number of components, once a sufficient
number of components have been included. Using the fact that different
principal components are orthogonal, we perform a PCA projection on
our descriptor space by whitening the descriptors and computing the
dot product with the principal components, keeping only the first $n$
components
\begin{equation}
\label{pcaprojection}
M(n)_{pi}^r = V^T(n)_{ij}\tilde{M}_{pj}^r \,\,\,\, ; \,\,\,\,  D(n)_i = V^T(n)_{ij}\tilde{D}_j.
\end{equation}
Here we indicate with $V^T(n)$ the truncation of $V^T$ to the first $n$
rows (i.e. $i$ can now range from $1$ to $n$). As described above, the
expectation is that most of the cosmological information is contained
in a small number of components $n<N_{b}$. We will describe in detail
below the choice we make for $n$, together with the sensitivity of our
results to this choice.

Looking at PCA from a geometrical perspective, the dimensionality
reduction problem is equivalent to the accurate reconstruction of the
coordinate chart of the descriptor manifold. As outlined in
ref.~\citep{astroMLText}, the coordinate chart constructed with the
PCA projection in eq.~(\ref{pcaprojection}) is accurate for reasonably
flat descriptor manifolds. When curvature becomes important, more
advanced projection techniques (such as Locally Linear Embedding) have
to be employed.  We postpone an investigation of such improvements to
future work.

\section{Results}
\label{results}

This section describes our main results, and is organized as follows.
We begin by showing the cosmological constraints from the CFHTLenS
data for the triplet $(\Omega_m,\sigma_8,w)$, as well as for an
alternative parameterization, $(\Omega_m,\Sigma_8,w)$, with
$\Sigma_8(\alpha)\equiv \sigma_8(\Omega_m/0.27)^\alpha$. In the next section we give a justification on why we fix $\alpha$ to a value of $0.55$. 
We then use our simulations to perform a robustness analysis of the parameter
confidence intervals with respect to the number of PCA components used
in the projection.
We finally study whether the constraints can be tightened by combining
different descriptors. A summary with the complete set of results,
along with the relevant Figures, is shown in Table \ref{summarytable}.

\subsection{Cosmological constraints}
\begin{figure*}
\begin{center}
\includegraphics[scale=0.45]{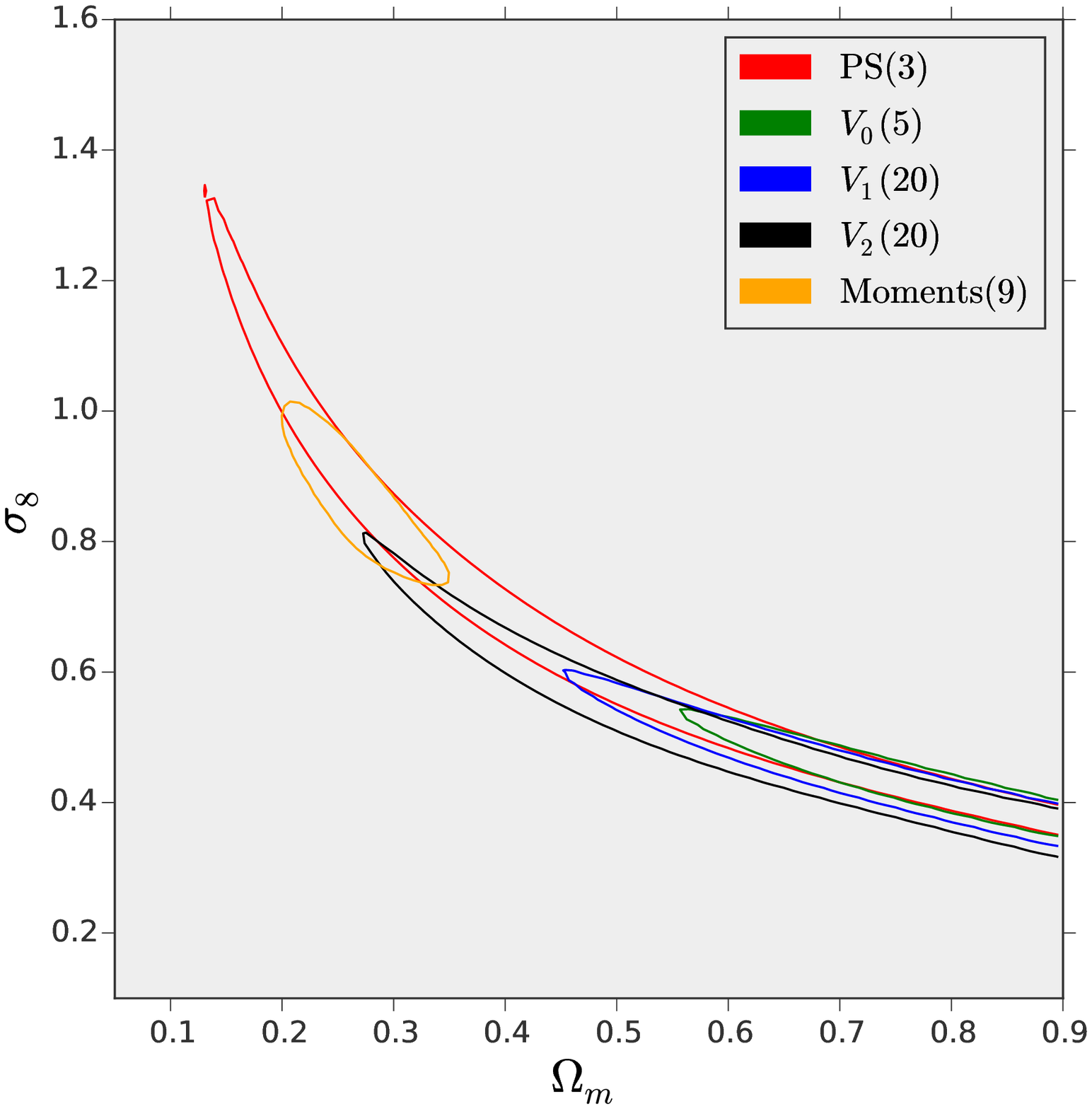}
\includegraphics[scale=0.45]{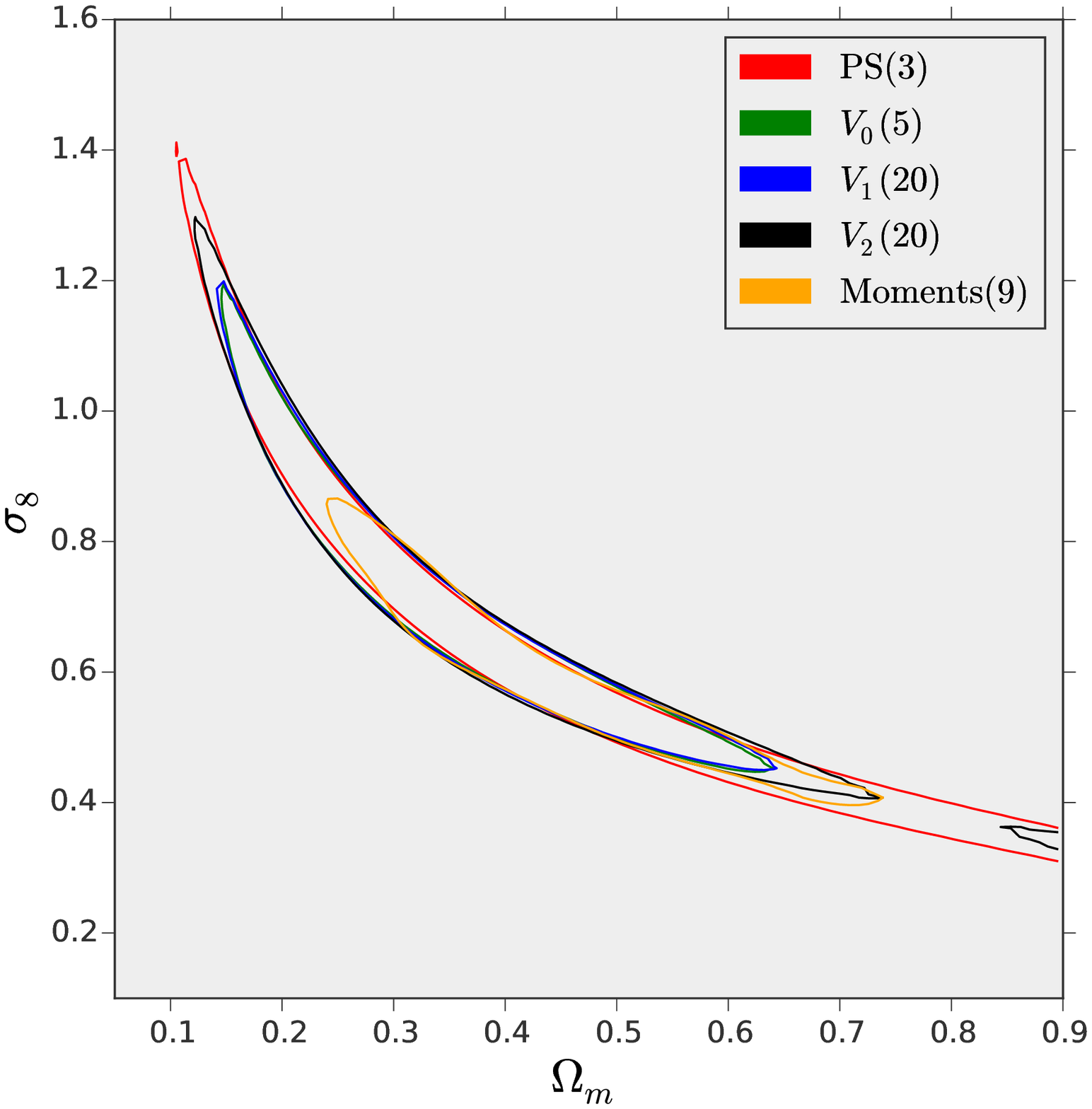}
\end{center}
\caption{$1\sigma$ (68\% CL) constraints on the $(\Omega_m,\sigma_8)$
  parameter doublet using the power spectrum (red), the three
  Minkowski functionals ($V_0$: green, $V_1$: blue, $V_2$: black) and
  the moments (orange). We show the constraints from the data (left
  panel) and from a mock observation constructed using the mean of
  1000 realizations in the \texttt{CFHTcov} simulation suite (right
  panel). The contours are calculated from the parameter likelihood
  function $\mathcal{L}$ marginalized over $w$. The parentheses near
  the descriptor label refer to the number of principal components
  included.}
\label{contours3single}
\end{figure*}

\begin{figure*}
\begin{center}
\includegraphics[scale=0.45]{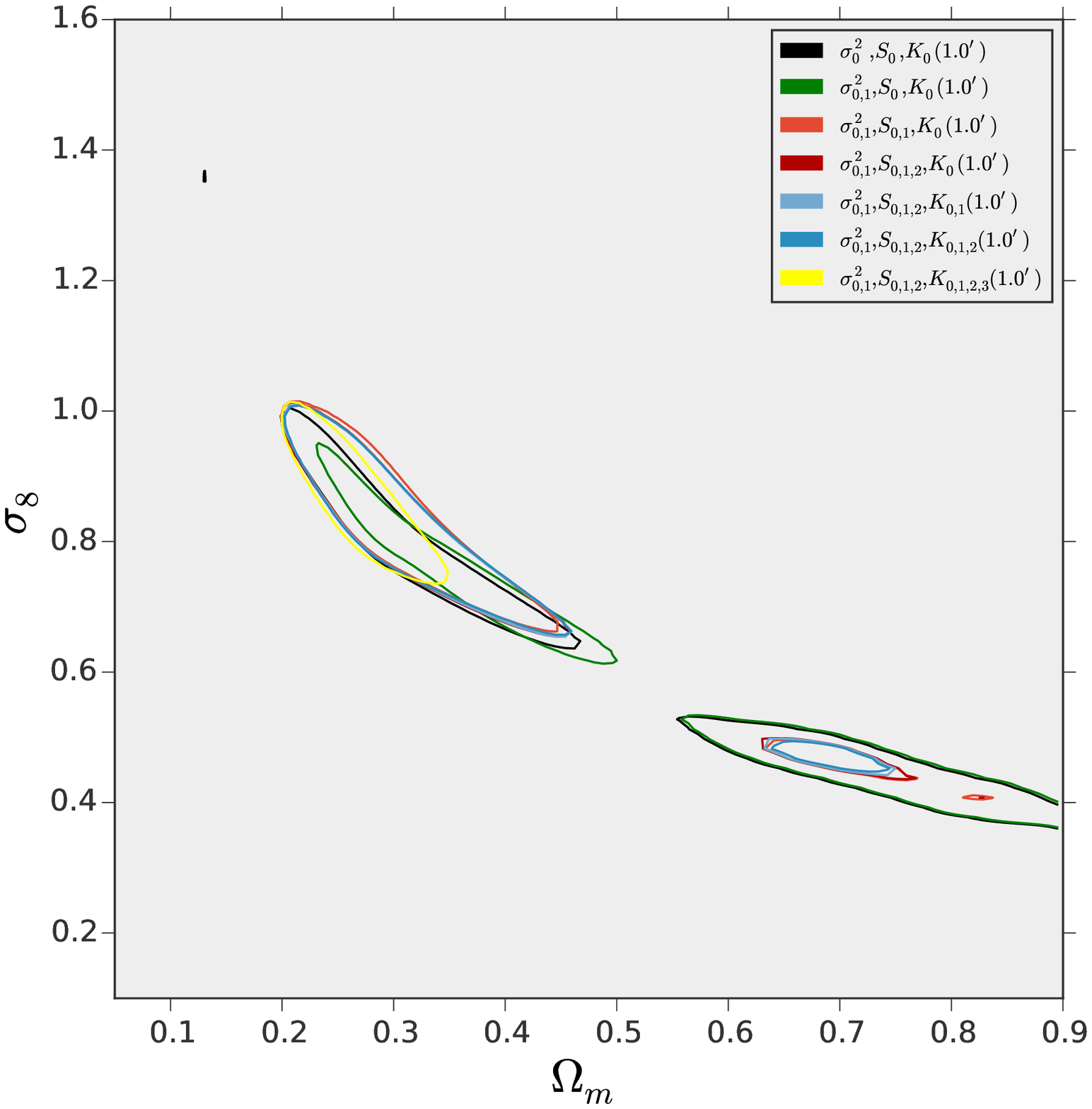}
\includegraphics[scale=0.45]{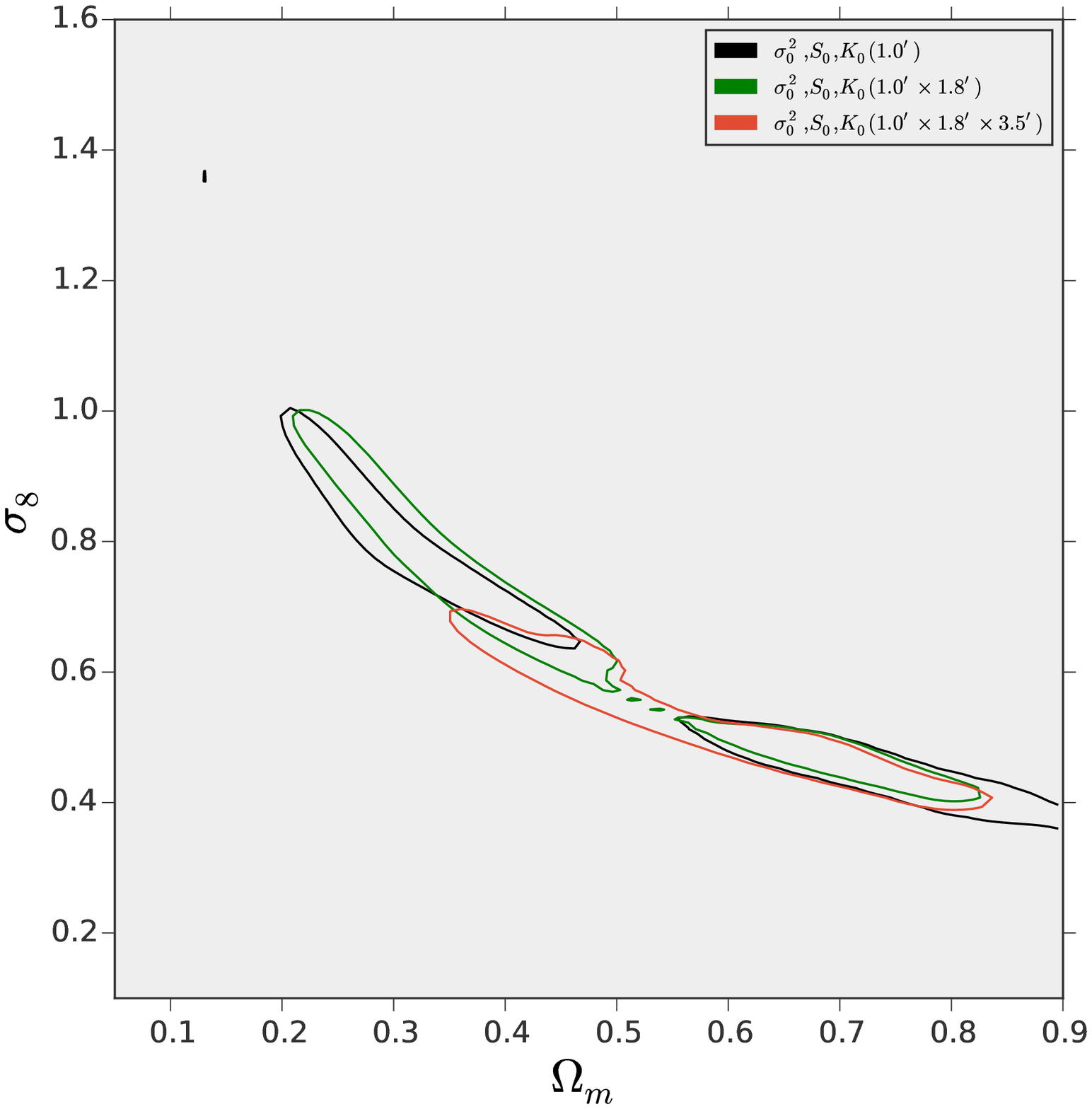}
\end{center}
\caption{$1\sigma$ (68\% CL) constraints on the $(\Omega_m,\sigma_8)$
  parameter doublet using moments, with different colors corresponding
  to different moment combinations (see eq.~\ref{momentestimator} for
  their definitions). We show the results from the one--point moments
  $\sigma_0^2,S_0,K_0$ (black curves; both left and right panels). In
  the left panel, we also show constraints obtained adding moments of
  gradients to the one--point moments. In the right panel, we combine
  one--point moments measured at different smoothing scales. }
\label{contoursMoments}
\end{figure*}

\begin{figure}
\begin{center}
\includegraphics[scale=0.45]{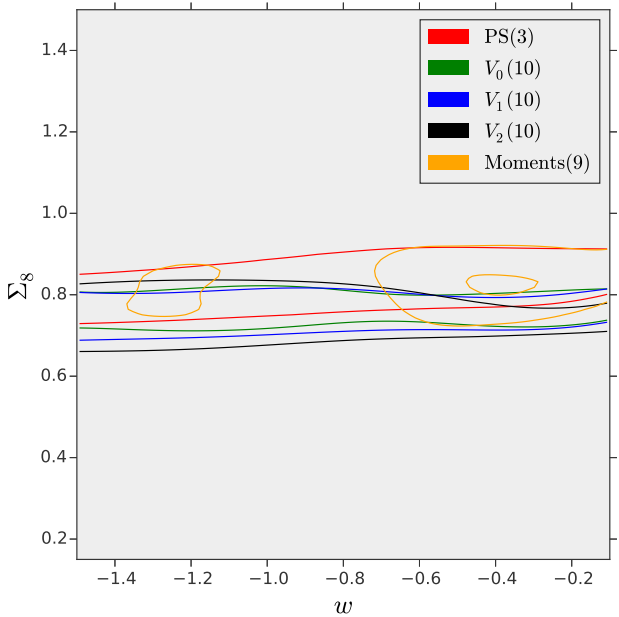}
\end{center}
\caption{$1\sigma$ (68\% C.L.) constraints on the $(w,\Sigma_8)$
  parameter doublet from the CFHTLenS data, obtained with the power
  spectrum (red), the three Minkowski functionals ($V_0$: green,
  $V_1$: blue, $V_2$: black) and the moments (orange).
   The contours are calculated from the parameter likelihood function
   $\mathcal{L}$ marginalized over $\Omega_m$, and the parentheses
   near the descriptor label refer to the number of principal
   components.}
\label{contours3singleRep}
\end{figure}

We first make use of equations
(\ref{parameterlikelihood})-(\ref{ennesigma}) to compute the $1\sigma$
constraints on cosmological parameters, using the triplet
$(\Omega_m,\sigma_8,w)$.  Figures \ref{contours3single} shows the
constraints in the $(\Omega_m,\sigma_8)$ plane, marginalized over $w$,
for both the CFHTLenS data, as well as from the mock data in our
simulations.  In Figure~\ref{contoursMoments}, we examine constraints
from different sets of moments, as well as using different smoothing
scales.  Figure~\ref{contours3singleRep} shows the confidence contours
in the $(w,\Sigma_8)$ plane, marginalized over $\Omega_m$. As this
figure shows, and as discussed further below, no meaningful
constraints were found on $w$ from CFHTLenS alone.

Because of the relatively small size of this survey, degeneracies
among the parameters can have undesirable effects on the constraints.
The well-known strong degeneracy between $\Omega_m$ and $\sigma_8$ is
evident in the long ``banana'' shaped contours in Figures
\ref{contours3single} and \ref{contoursMoments}.  To mitigate the
effect of this degeneracy, in addition to the usual triplet
$(\Omega_m,\sigma_8,w)$, we consider an alternative parameterization,
built with the triplet $(\alpha,\Sigma_8,w)$ where $\alpha$ is a
constant, and
$\Sigma_8(\alpha)\equiv\sigma_8(\Omega_m/0.27)^\alpha$. While
$\Omega_m$ and $\sigma_8$ are poorly constrained due to degeneracies,
the $\Sigma_8(\alpha)$ combination lies in the direction perpendicular
to the error ``banana'' at the pivot point $\Omega_m=0.27$.  
This is the direction of the lowest variance
$\mathcal{L}(\Omega_m,\sigma_8)$ for a suitable choice of $\alpha$,
and hence has a much smaller relative uncertainty. We can derive the
optimal value of $\alpha$ from the full three dimensional likelihood
$\mathcal{L}(\Omega_m,w,\sigma_8)$, from which we can compute the
expectation values
\begin{equation}
\mathds{E}(\alpha) = \langle\Sigma_8(\alpha)\rangle \,\,\, ; \,\,\, \mathds{V}(\alpha) = \langle(\Sigma_8(\alpha)-\mathds{E}(\alpha))^2\rangle
\end{equation}
and minimizing the ratio $\sqrt{\mathds{V}}/\mathds{E}$ with respect
to $\alpha$. The expectation values are taken over the entire parameter box. 
This procedure yields a value $\alpha\approx0.55$ for the
statistical descriptors that we consider, 
consistent with what is found in the literature (see \citep{CFHTKilbinger} for example). Although $\alpha$ can mildly depend on the type of descriptor considered, we choose to keep it fixed, knowing that the width of the $\Sigma_8$ likelihood cannot vary significantly with different choices of $\alpha$.  
We show the probability distribution of the best-constrained parameter
$\Sigma_8$ (marginalized over $\Omega_m$ and $w$) in Figure
\ref{likelihoodSi8single}. 

We discuss the results of this section in \S~\ref{discussion} below.

\subsection{Robustness}

The cosmological constraints should in principle be insensitive to
$N_b$, once a sufficient number of bins are used, but inaccuracies in
the covariance (due to a limited number of realizations) can introduce
an $N_b$ dependence. Our binning choices are summarized in Table \ref{desctable}.  
Here we show that the cosmological constraints
derived in this paper are numerically robust, i.e. they are reasonably
stable, once we consider a large enough number $n$ of Principal
Components. 

\begin{figure*}
\includegraphics[scale=0.4]{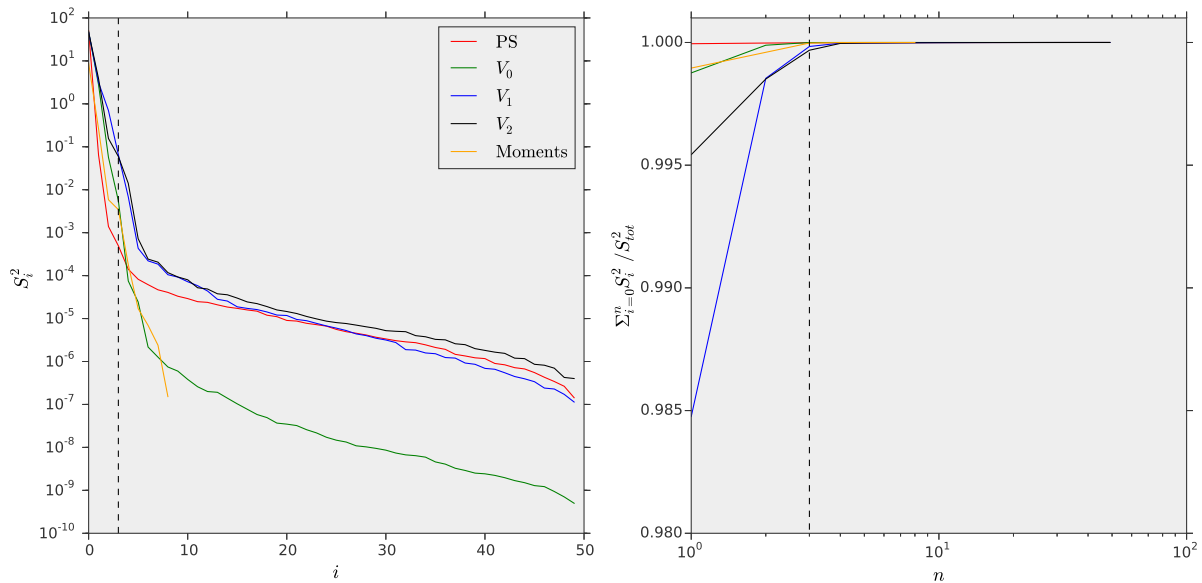}
\caption{Results from the Principal Components Analysis (PCA) of the
  binned power spectrum (red), the three Minkowski functionals ($V_0$:
  green, $V_1$: blue, $V_2$: black) and the moments (orange).
  The left panel shows the magnitudes of the PCA eigenvalues $S_i^2$
  and the right panel shows their cumulative sum, normalized to
  unity. A dashed vertical (black) line has marks $n=3$ for
  reference.}
\label{pcafig}
\end{figure*}

\begin{figure*}
\includegraphics[scale=0.4]{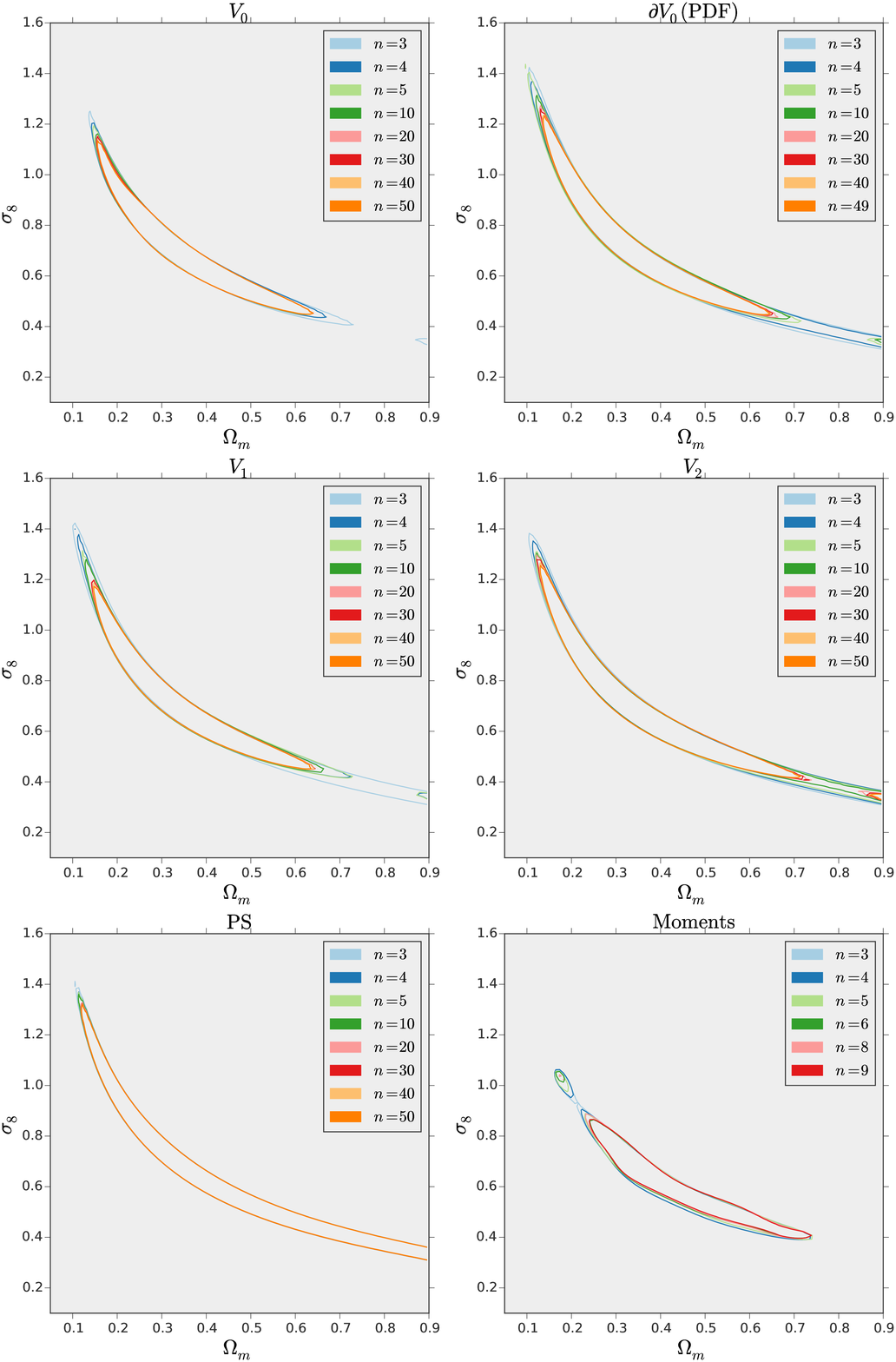}
\caption{The dependence of the $1\sigma$ contours in the
  $(\Omega_m,\sigma_8)$ plane on the number of PCA components,
  obtained from a mock observation constructed with the
  \texttt{CFHTcov} simulations. The different panels refer to the
  different descriptors (from left to right, top to bottom) $V_0$,
  $\partial V_0$(PDF), $V_1$, $V_2$, power spectrum and moments.  The
  labels in each panel show the number of PCA components included to
  obtain contour with different colors.}
\label{robustnessfig}
\end{figure*}

Figure~\ref{pcafig} shows the PCA eigenvalues from the SVD
decomposition of our binned descriptor spaces (following the
discussion in \S~\ref{pcasection}), as well as the cumulative sum of
these eigenvalues, normalized to unity. Figure \ref{robustnessfig}
shows the dependence of the $(\Omega_m,\sigma_8)$ constraints on the
number of principal components $n$.  

These figures clearly indicate that we only need a limited number of
components in order to capture the cosmological information contained
in our descriptors. The eigenvalues diminish rapidly with $n$, and, in
particular, the confidence contours converge to good ($\lesssim 10\%$)
accuracy typically for $n=5-10$ (depending on the descriptor).
This finding also addresses the inaccuracy of the MF emulator at high
thresholds, pointed out in Figure~\ref{emulatorAccuracy}. By keeping a
limited number of principal components, we are able to prevent the
inaccurate high--threshold bins, which have a low constraining power,
from contributing to the parameter confidence levels.

\begin{figure}
\begin{center}
\includegraphics[scale=0.45]{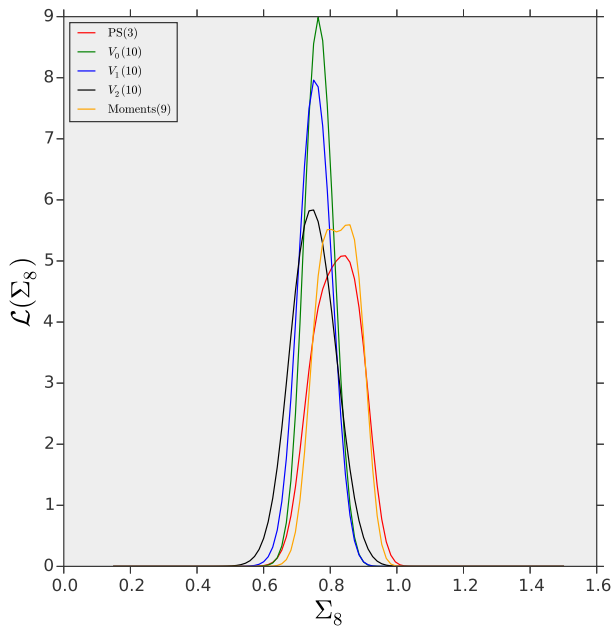}
\end{center}
\caption{The likelihood of the best-constrained parameter combination
  $\Sigma_8(\alpha)\equiv\sigma_8(\Omega_m/0.27)^\alpha$ from the
  CFHTLenS data, obtained with the power spectrum (red), the three
  Minkowski functionals ($V_0$: green, $V_1$: blue, $V_2$: black) and
  the moments (orange).
  The likelihood was computed with a constant optimized $\alpha=0.55$,
  but marginalized over both $\Omega_m$ and $w$. 
  The parentheses near the descriptor label refer to the number of
  principal components.}
\label{likelihoodSi8single}
\end{figure}

\begin{table*}
\begin{tabular}{c|c|c||c}
Parameters & Descriptors & Short description & Relevant Figures \\ \hline \hline
$(\Omega_m,\sigma_8)$ & PS(3),$V_0(5),V_1(20),V_2(20)$,LM(9) &\pbox{20cm}{$1\sigma$ constraints from CFHTLenS \\ and mock observations}  & \ref{contours3single},\ref{contours3single}b \\ \hline
$(\Omega_m,\sigma_8)$ & $(\sigma_i^2,S_i,K_i)$ & \pbox{20cm}{$1\sigma$ constraints from CFHTLenS \\ using $\kappa$ moments \\ combined at different $\theta_G$}  & \ref{contoursMoments},\ref{contoursMoments}b \\ \hline
$(w,\Sigma_8)$ & PS(3),$V_0(10),V_1(10),V_2(10)$,LM(9) & contours from CFHTLenS & \ref{contours3singleRep} \\ \hline 
$\Sigma_8$ & PS(3),$V_0(10),V_1(10),V_2(10)$,LM(9) & $\mathcal{L}(\Sigma_8)$ from CFHTLenS & \ref{likelihoodSi8single} \\ \hline
- & PS,$V_0,V_1,V_2$,LM & PCA eigenvalues  & \ref{pcafig} \\ \hline
$(\Omega_m,\sigma_8)$ & PS,$V_0,V_1,V_2$,LM & Stability of contours & \ref{robustnessfig} \\ \hline 
$(\Omega_m,\sigma_8)$ & PS(3)$\times V_0(5)\times V_1(20)\times V_2(20)\times$LM(9) & \pbox{20cm}{constraints from CFHTLenS \\ combining statistics} & \ref{contours3combined} \\ \hline
$(w,\Sigma_8)$ & PS(3)$\times V_0(10)\times V_1(10)\times V_2(10)\times$LM(9) & \pbox{20cm}{constraints from CFHTLenS \\ combining statistics} & \ref{contours3combined}b \\ \hline 
$\Sigma_8$ & PS(3)$\times V_0(10)\times V_1(10)\times V_2(10)\times$LM(9) & \pbox{20cm}{$\mathcal{L}(\Sigma_8)$ from CFHTLenS \\ combining statistics} & \ref{likelihoodSi8cross} \\ \hline
\end{tabular}
\caption{Summary of our results and related figures.}
\label{summarytable}
\end{table*}


\begin{figure*}
\begin{center}
\includegraphics[scale=0.45]{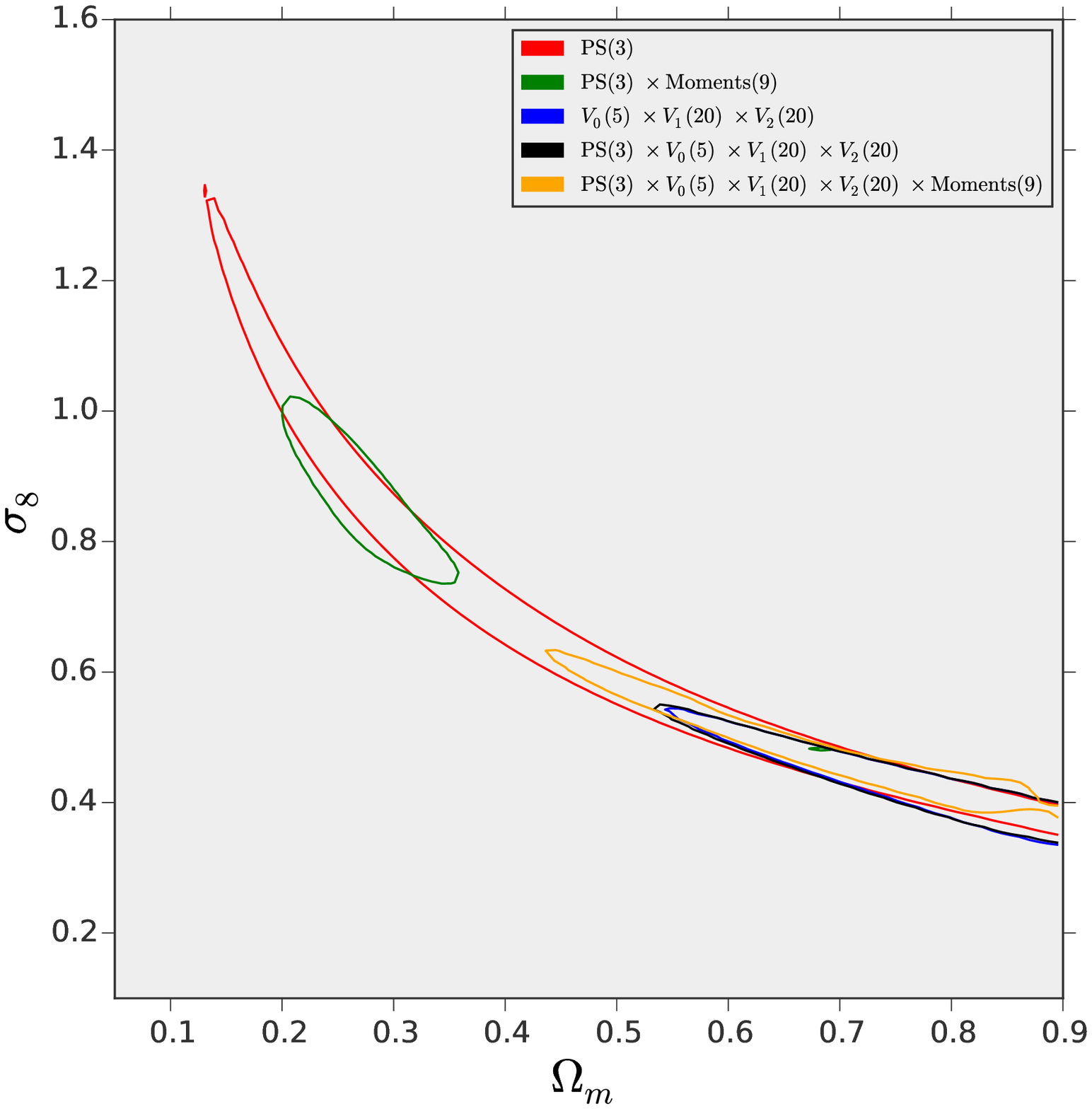}
\includegraphics[scale=0.45]{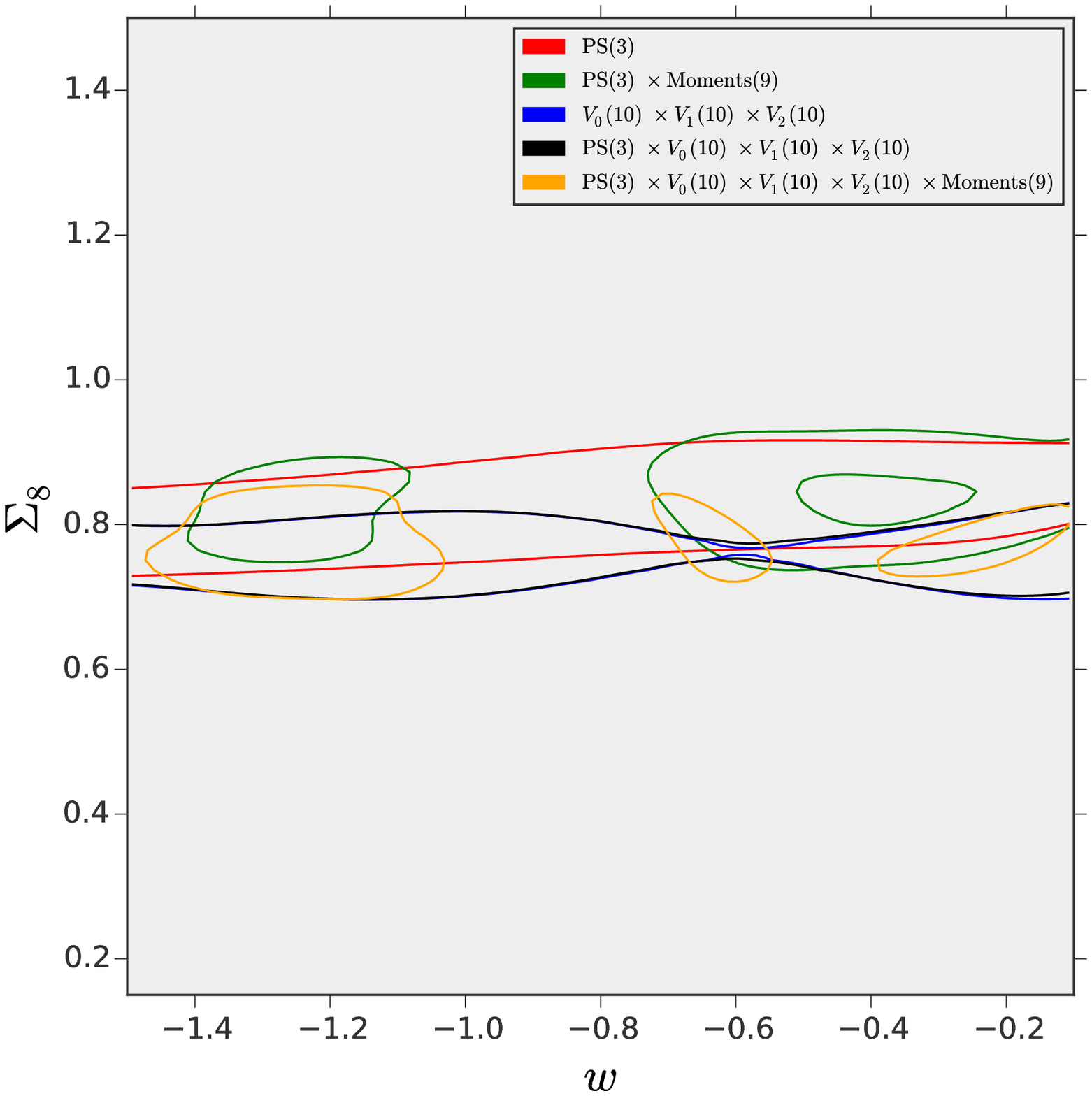}
\end{center}
\caption{$1\sigma$ constraints on the $(\Omega_m,\sigma_8)$ (left
  panel) and $(w,\Sigma_8)$ (right panel) doublets, using the power
  spectrum (PS) alone (red), the MFs alone (blue), as well as using
  different combinations of descriptors: PS$\times$Moments (green),
  PS$\times$MFs (black) and PS$\times$MFs$\times$Moments (orange). The
  likelihood function has been marginalized over $w$ (left panel) and
  $\Omega_m$ (right panel). The parentheses next to each descriptor
  label refers to the number of PCA components included.}
\label{contours3combined}
\end{figure*}

\begin{figure}
\begin{center}
\includegraphics[scale=0.45]{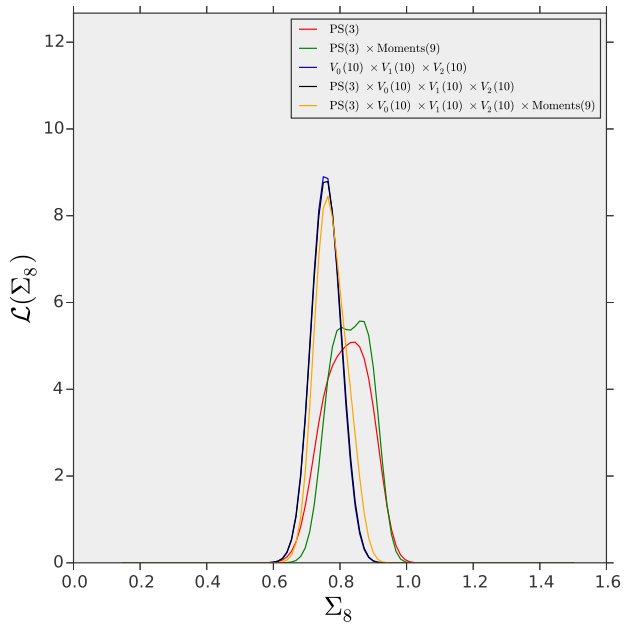}
\end{center}
\caption{The probability distribution of the best-constrained
  parameter $\Sigma_8$ from the CFHTLenS data, using the power
  spectrum (PS) alone (red), the MFs alone (blue), as well as using
  different combinations of descriptors: PS$\times$Moments (green),
  PS$\times$MFs (black) and PS$\times$MFs$\times$Moments (orange).
  The likelihood function has been marginalized over $\Omega_m$ and
  $w$. The parentheses next to each descriptor label refers to the
  number of PCA components.}
\label{likelihoodSi8cross}
\end{figure}

\begin{table*}
\begin{tabular}{c|c}
Descriptor & $\Sigma_8=\sigma_8\Omega_m^{0.55}$ \\ \hline \hline
PS(3) & $0.84^{+0.06}_{-0.09}$\\
PS(3) $\times$ Moments(9) & $0.86^{+0.02}_{-0.09}$ \\
$V_0(10)\times V_1(10) \times V_2(10)$  & $0.75^{+0.07}_{-0.04}$ \\
PS(3) $\times V_0(10)\times V_1(10) \times V_2(10)$ & $0.76^{+0.04}_{-0.05}$ \\
PS(3) $\times V_0(10)\times V_1(10) \times V_2(10) \times$ Moments(9) & $0.76^{+0.06}_{-0.04}$ \\ \hline
\end{tabular}
\caption{Tabulated values of $1\sigma$ constraints on $\Sigma_8$ corresponding to Figure \ref{likelihoodSi8cross}}
\label{Sigma8Table}
\end{table*}

\subsection{Combining statistics}

Different statistics can include complementary cosmological
information, allowing their combinations to tighten the
constraints. Previous work using multiple lensing descriptors in
CFHTLenS alone included combining the power spectrum and peak
counts~\citep{Companion}, combining the power spectrum and Minkowski functionals ~\citep{CFHTMasato} ,and combining quadratic (2PCF) statistics
with cubic statistics derived from the 3PCF of the CFHTLenS $\kappa$ field~\citep{CFHTFu}.

The procedure we adopt here is as follows. Consider two binned
descriptors, $d_{1,i},d_{2,j}$ where the indices $i,j$ correspond to
bin numbers.  We first compute each single--descriptor constraint as a
function of the number of PCA components, as in Figure
\ref{robustnessfig}. We then determine the minimum number of PCA
components $n_{1,2}$ needed for the constraints to be stable. We next
construct the vector $d_{1\times2} = \{d_1(n_1),d_2(n_2)\}$ and
consider this as the combined ($n_1+n_2$)--dimensional descriptor
vector. This procedure naturally allows us to account for the
cross--covariance between different binned descriptors. An analogous
procedure can be used to combine multiple (three or more)
descriptors.

We show constraints from different descriptor combinations in the
$(\Omega_m,\sigma_8)$ and $(w,\Sigma_8)$ planes in Figure
\ref{contours3combined}, and on the best-constrained parameter
$\Sigma_8$ in Figure \ref{likelihoodSi8cross}. We also provide a tabulated version of the $\Sigma_8$ constraints ($1\sigma$) in Table \ref{Sigma8Table}. We discuss these findings in the next section.

\section{Discussion}
\label{discussion}

In this section we discuss the results shown in \S~\ref{results}
above, with particular focus on the constraints on cosmology.

As pointed out in \S~\ref{pcasection}, the choice of the number of
bins, $N_b$, is an important issue. In order to ensure that our
results are insensitive to $N_b$, we adopted a PCA projection
technique to reduce the dimensionality of our descriptor spaces. The
left panel of Figure~\ref{pcafig} shows that the PCA eigenvalues for
all of our descriptors decrease by about 4 orders of magnitude from
$n=1$ to $n=3$. The right panel of this figure shows that more than
99\% of the descriptor variances are captured by including only the
first $n=3$ components.

This does not necessarily mean, however, that the cosmological
information is captured by the first 3 PCA components: in principle,
one of the higher-$n$ PCA components could have an unusually strong
cosmology-dependence, and could impact the confidence levels.  To
address this possibility, we determined the $1\sigma$ contour sizes as
a function of $n$ in Figure~\ref{robustnessfig}.  This figure shows
that the first 3 components indeed capture essentially all the
information contained in the power spectrum. However, this is not true
for the other descriptors. In particular, we find that $n\geq5$
components are necessary for $V_0$, and $n\geq20$ components for $V_1$
and $V_2$, in order for the $(\Omega_m,\sigma_8)$ contours to be
stable at the $\sim$5\% level. All nine moments need to be included
for the moments contours to be stable to this accuracy.

These results are slightly different when we study the $(w,\Sigma_8)$
constraints (with $\alpha$ fixed at $\alpha=0.55$ as discussed
above). In this case we find that the optimal choice for all three MFs
is $n=10$, while the number of components required for the PS and
Moments remain at $n=3$ and $n=9$, respectively. (These results are
not shown, but obtained analogously to the Figures above.)

We now discuss the main scientific findings of this work. In
Figure~\ref{contours3single}, we show the $1\sigma$ constraints on the
$(\Omega_m,\sigma_8)$ doublet from the CFHTLenS data. The MF
constraints appear to be biased towards the low--$\sigma_8$,
high--$\Omega_m$ region. Here and throughout the remainder of this
paper, by "biased" (or "unbiased") we refer to being incompatible (or
compatible) with the concordance fiducial values at $1\sigma$ obtained
in other experiments. For example, the current best-fit values of
$(\Omega_m,\sigma_8)=(0.32,0.83),(0.28,0.82)$ from cosmic microwave background
anisotropies measured respectively by the Planck
~\citep{PlanckXVI2013} and WMAP~\citep{WMAP9} satellites lie beyond the 99\% likelihood
contours obtained from the three MFs (not shown).

This discrepancy may be due to uncorrected systematics in the CFHTLenS
data, amplified by the $(\Omega_m,\sigma_8)$ degeneracy.  As a test of
our analysis pipeline, when we try to constrain mock observations
based on simulations (shown in the
right panel of Figure \ref{contours3single}), we recover the correct
input position of the $1\sigma$ contours.  It is important to note,
however, that the mock observations to which the right panel of Figure
\ref{contours3single} refers, were built with the mean of $R=1000$
realizations of the \texttt{CFHTcov} simulations. We found that it is
possible to find some rare (10 out of the 1,000) realizations for
which the best fit for $(\Omega_m,\sigma_8)$ lies in the lower right
corner, near the location of the best-fit from the data. While this
could provide an alternative explanation of the bias from the MFs, the
likelihood of this happening is very small ($\lesssim1\%$).

We observe that the moments give the tightest constraint on
$(\Omega_m,\sigma_8)$.  Furthermore, this constraint is unbiased, in
the sense defined above: it includes the current concordance values
for these parameters within 1$\sigma$.  This leads us to conclude that
the bias in the constraints from the MFs is due to systematic errors,
rather than the rare statistical fluctuations found above.
The fact that the moments are useful for deriving unbiased
cosmological constraints has been noted in previous work, which
examined the biases caused by spurious shear
errors~\citep{PetriSpurious}.

In order to determine the origin of the tight bounds derived from
moments, we studied the contribution of each individual moment to the
constraints. Figure \ref{contoursMoments} shows the evolution of the
$(\Omega_m,\sigma_8)$ constraints as we add increasingly higher-order
moments to the descriptor set. Since we are constraining 3
cosmological parameters, we start by considering the set of the three
traditional one--point moments which do not involve gradients,
i.e. the variance, skewness, and kurtosis $(\sigma_0^2,S_0,K_0)$. We
then add the remaining six moments of derivatives one by one, starting
from the quadratic moments. 

Figure \ref{contoursMoments} shows that the biggest improvement on the
parameter bounds comes from including quartic moments of derivatives
(i.e. $K_i$ with $i\ge1$) in the descriptor set. This might explain
why \citep{CFHTFu} find only relatively weak contour tightening
($\sim10\%$) when adding three--point correlations to quadratic
statistics, since the main improvement comes from higher moments of
$\kappa$ derivatives. Ref.~\citep{CFHTFu} consider one--point,
third--order moments, combined for multiple smoothing scales. Figure
\ref{contoursMoments} explicitly shows, however, that smoothing scale
combinations are not as effective as moments of derivatives in
constraining the $(\Omega_m,\sigma_8)$ doublet. Our results agree with
an early prediction \citep{moments4} that the kurtosis of the shear
field can help in breaking degeneracies between $\Omega_m$ and
$\sigma_8$.  Here we found that considering quartic moments of
gradients further helps in breaking this degeneracy.

As noted above, the bias in the $(\Omega_m,\sigma_8)$ constraints is
amplified by the cosmological degeneracy of these parameters. To
mitigate this effect, we consider the combination of $\Omega_m$ and
$\sigma_8$ that lies orthogonal to the most degenerate direction,
namely $\Sigma_8=\sigma_8(\Omega_m/0.27)^{0.55}$. Figure
\ref{contours3singleRep} shows the $1\sigma$ constraints for the
$(w,\Sigma_8)$ doublet, while Figure \ref{likelihoodSi8single} shows
the marginalized $\Sigma_8$ likelihood from the CFHTLenS data. The
CFHTLenS survey constrains the $\Sigma_8$ combination to a value of
$\Sigma_8=0.75\pm0.04(1\sigma)$ using the full descriptor set, in
agreement with values previously published by the CFHTLenS
collaboration \citep{CFHTKilbinger}. 

These figures also show that the current dataset is insufficient to
constrain $w$ to a reasonable precision.
This is consistent with the previous analyses of
CFHTLenS~\citep{CFHTKilbinger,CFHTFu,Companion,CFHTMasato}.  We also
note that ref.~\citep{CFHTMasato} obtained the best-fit value of
$w\approx -2$ (but with large errors that include $w=-1$ at
1$\sigma$).  We found a similar result when using a Fisher matrix to
compute confidence levels. Since the Fisher matrix formalism is equivalent to a linear approximation of our emulator (in which all cosmological parameter dependencies are assumed to be linear), we thus attribute this bias to the oversimplifying assumption of linear cosmology-parameter dependence of the descriptors.  
Although the right panel in Figure
\ref{contours3combined} shows that the moments confine $w$ to isolated
regions in parameter space, we note that $w=-1$, the value favored by other existing experiments, is excluded at the $1\sigma$
level. The $2\sigma$ contours (not shown in the figure) join, and
include $w=-1$.

Regarding the parameter biases, our results overall are in accordance
with \citep{PetriSpurious}, namely, that unaccounted systematics
result in larger parameter biases when the constraints are derived
from the MFs, and that the LM statistic is less biased. However, for
the CFHTLenS data the MFs can still effectively constrain the
non-degenerate direction in parameter space, $\Sigma_8$ (Figure
\ref{likelihoodSi8single}).

Finally, we studied whether the combination of different statistical
descriptors can help in tightening the cosmological constraints. We
show the effects of some of these combinations in Figures
\ref{contours3combined} and \ref{likelihoodSi8cross}. The left panel
of Figure \ref{contours3combined} shows that, although combining the
power spectrum and the moments with the Minkowski functionals helps
tighten the $(\Omega_m,\sigma_8)$ constraints, it does not help in
reducing the inherent parameter bias of the MFs. The right panel of Figure
\ref{contours3combined} shows that even with these statistics
combined, $w$ remains essentially unconstrained. \citep{CFHTHeymansTomo} found that even weak lensing tomography alone is unable to constrain $w$ sensibly. 
Figure \ref{likelihoodSi8cross} shows that the $\Sigma_8$ combination is
already well constrained by any of the descriptors alone, without the
need of combining different descriptors.
This further clarifies that the non--quadratic descriptors mainly help
to break degeneracies, tightening contours along the
degenerate direction.

\section{Conclusions}

In this final section we summarize the main conclusions of this work:

\begin{itemize}

\item We find that the power spectrum, combined with the moments of
  the $\kappa$ field provides the tightest constraint on the
  $(\Omega_m,\sigma_8)$ doublet from the CFHTLenS survey data. The
  tightness of these constraints comes mainly from the
  moments. Evidence of the unbiased nature of constraints from the
  moments has been found in \citep{PetriSpurious}. We further find
  that the largest improvement on parameter bounds is achieved when we
  include the quartic moments of derivatives in the descriptor set. This
  level of improvement cannot be achieved by combining one-point
  moments at different smoothing scales.

\item Although weak lensing surveys are a promising technique to
  constrain the DE equation of state parameter $w$, reasonable
  constraints cannot be obtained with the CFHTLenS survey alone, even
  when using additional sets of descriptors that go beyond the
  standard quadratic statistics.

\item When studying the cosmological information contained in the
  CFHTLenS data, special attention must be paid to the effect of
  residual systematic biases. While these residual systematics are
  found to be unimportant when constraining cosmology with the power
  spectrum alone, we find that these systematics need to be corrected
  to obtain unbiased constraints on the $(\Omega_m,\sigma_8)$ doublet
  using the Minkowski functionals. We are aware that, when trying to explain the discrepancy between weak lensing and CMB constraints using the Minkowski functionals, there might be other effects to be considered, namely non--Gaussian error correlations in the descriptors and inaccuracies of the simulations on small scales. These inaccuracies could in principle affect the excursion set reconstruction at high $\kappa_0$ thresholds. We will investigate these additional sources of error in future work.

\item For the CFHTLenS data set, Minkowski functionals can effectively
  constrain the non-degenerate direction in parameter space,
  $\Sigma_8$, where the amplifying effects of degeneracy are
  mitigated.The Minkowski functionals alone are sufficient to
  constrain the $\Sigma_8$ combination to a value of
  $\Sigma_8=0.75\pm0.04$ at $1\sigma$ significance level. This agrees
  with the value previously published by the CFHTLenS collaboration
  within $1\sigma$. Some tensions with Planck \citep{PlanckXVI2013}
  still remain.

\end{itemize}

Possible future extensions of this work include simulating
higher--dimensional parameter spaces (including, for example, the
Hubble constant $H_0$, and allowing a tilt in the power spectrum or a
time-dependence of the DE equation of state $w$), and combining the
CFHTLenS constraints with different cosmological probes from
large-scales structures and the CMB. The latter can help in breaking
the $\Omega_m,\sigma_8$ degeneracy, and allow improvements in the the
constraints on $w$.  The techniques developed here can be applied to
larger, soon-forthcoming survey data sets, such as the Dark Energy Survey (DES)~\citep{DES}, Subaru~\citep{Subaru}, WFIRST~\citep{WFIRST} and LSST~\citep{LSST}.

\section*{Acknowledgements}

The simulations in this work were performed at the NSF Extreme Science
and Engineering Discovery Environment (XSEDE), supported by grant
number ACI-1053575, at the Yeti computing cluster at Columbia
University, and at the New York Center for Computational Sciences, a cooperative
effort between Brookhaven National Laboratory and Stony Brook
University, supported in part by the State of New York. This work was
supported in part by the U.S. Department of Energy under Contract
Nos. DE-AC02-98CH10886 and DE-SC0012704, and by the NSF Grant
No. AST-1210877 (to Z.H.) and by the Research Opportunities and
Approaches to Data Science (ROADS) program at the Institute for Data
Sciences and Engineering at Columbia University.

\bibliography{ref}
\label{lastpage}
\end{document}